\documentstyle[epsfig,aps]{revtex}
\def\bea{\begin{eqnarray}}
\def\eea{\end{eqnarray}}
\def\kdotr {{\bf k_t}\cdot{\bf r_t}}
\begin{document}
\draft
\title{\hskip4in  NT@UW-99-59
\\
Coherent QCD phenomena in the
Coherent Pion-Nucleon and Pion-Nucleus
Production of Two Jets at High Relative Momenta}
\author{L. Frankfurt}
\address{ School of Physics and Astronomy, \\
Tel Aviv University, 69978
Tel Aviv, Israel}
\author{G. A. Miller}
\address{Department of Physics, Box 351560\\
University of Washington \\
Seattle, WA 98195-1560, U.S.A.}
\author{M. Strikman}
\address{ Department of Physics,
Pennsylvania State University,\\
University Park, PA  16802, USA}
\maketitle
\begin{abstract}
We use  QCD  to compute the cross section for high-energy coherent
production of a dijet (treated as a $q\bar q$ moving at high relative
transverse momentum, $\kappa_t $) from a nucleon and a nuclear target.
The direct evaluation of the relevant Feynman diagrams shows
that, in the target rest frame, the space-time evolution of this reaction
is dominated by the process  in which the high
$\kappa_t$ $q\bar q$ component (point like configuration) of
the pion wave function is formed before reaching the target.
This point like configuration then interacts through two gluon
exchange with the target. In the approximation of keeping the leading
order in powers of  $\alpha_s$ and in the leading log approximation in
$\alpha_{s}\ln (\kappa_{t}^2/\Lambda_{QCD}^2), $
the amplitudes for other processes are shown to be
smaller by  at least a power of $\alpha_{s}$
and/or powers of Sudakov-type form factors  and the small probability, $w_2$,
to find a $q\bar q$ pair with no gluons at an average separation
between constituents.
Thus the high $\kappa_t$ component of pion wave function, including
the contribution of Gegenbauer polynomials of rank $n>0$, can be
measured in principle at sufficiently large  values of $\kappa_t^2$. At
large values of $\kappa_t^2$, the resulting dominant amplitude is
proportional to $z(1-z)\alpha_s(k_t^2)\kappa_t^{-4}
(\ln{\kappa_t^2\over \Lambda^2})^{C_F\over \beta}$
($z$ is the fraction light-cone (+) momentum carried by the quark in
the final state, $\beta$ is the coefficient in the running coupling constant)
times the skewed gluon distribution of the target.
For  pion scattering by a  nuclear target,  this means that
at fixed $x_{N}=2\kappa_{t}^2/s$  (but $\kappa_{t}^2\to \infty$ )
the nuclear process in which there is only a single
interaction is the most important one to contribute to the reaction. Thus
in this limit color  transparency  phenomena should
occur--initial and final state interaction effects
are absent for sufficiently large values of $\kappa_t$. These findings are in
accord with the recent experiment performed at FNAL. We  also
re-examine  a potentially important nuclear multiple scattering
correction which is positive and varies as the length of the nucleus
divided by an extra factor of $1/\kappa_t^4$. The meaning of the
signal obtained from the experimental measurement of pion diffraction
into two  jets is also critically examined and  significant corrections are
identified. We show also that  for  values of $\kappa_t$ achieved  at
fixed target energies, dijet production by the electromagnetic field of the
nucleus leads to an insignificant correction which gets more
important as $\kappa_t$ increases.
We explain also that the same regularities are valid for photo-production of
forward light quark dijets.
\end{abstract}

\section{Introduction}
The theory of strong interactions,  QCD, contains many specific predictions
regarding the space-time evolution of high energy coherent processes.
A review describing the many interesting qualitative results that have
been obtained in this rapidly developing field is provided in
Ref.~\cite{AC}. The aim of this paper is to use a specific example of
a completely calculable process to demonstrate the general properties
of space-time evolution of hard exclusive processes in QCD. In particular,
we consider  a process in which a high momentum ($\sim$ 500 GeV)  pion
undergoes a coherent interaction  with a nucleus in  such a way  that
the final state consists of two jets (JJ) (formed by a $q\bar q$ pair)
moving at high transverse relative momentum greater than about 2 or 3
GeV. The process of two-jet production was first discussed for both photon
and pion projectiles interacting with a nucleon target \cite{Randa}, and
Ref.~\cite {bb} introduced the possibility of using  this process to probe
the nuclear filtering of small color dipoles. Estimates of Ref.~\cite {bb}
found that for heavy nuclei,  nuclear filtering causes the exclusive
dijet production to decrease exponentially as $\kappa_t^2$ increases.
Hence an overall increase of the total diffractive cross
sections was suggested as a good signature of the nuclear filtering of
small size configurations\cite{comment}. In \cite{fms93} we presented
the first application of QCD to the process of dijet production at
large $\kappa_t$ by generalizing  QCD factorization theorems,
predicted a nuclear dependence which is qualitatively different from
that suggested in \cite{bb}, and a  $\kappa_t$ dependence which
differs   by a power of $\kappa_t^2$ from \cite{Randa} and qualitatively
from that discussed in \cite{bb}. We also argued that this process can
be used to directly measure the behavior of the $q\bar q$ component of the
pion's light cone wave function for large values of  $\kappa_{t}$.

If one wishes to describe hard diffractive processes, it is important
to realize  that the effective number of bare particles in the light
cone wave function of the projectile depends strongly on the
longitudinal distances involved.  If these  distances are small, and
the process involves high momentum transfer,  (as occurs in computing
hadron electromagnetic form factors) the main contribution
originates from the Fock component of the hadron wave function
containing the minimal number of constituents. On the contrary, if the
longitudinal distances  are sufficiently large, the  minimal Fock
component (a $q\bar q$ pair in our case) will develop additional
components such as $q\bar q$ pairs and  gluons. Thus for those
processes in which  infinite longitudinal distances are
involved, the number of partons in the light cone wave function of the 
projectile would always be infinite. For  processes initiated by a
spatially small colorless dipole, using the   QCD factorization
theorem allows one to trace the origin of wee partons (carrying a
small fraction of the momenta)  as arising from the space-time evolution
of the projectile's  minimal Fock component,  and  to
include these effects in the (skewed) parton distribution of the target.

We now discuss the basic physics of the dijet production process.
The selection of the final state to be  a $q\bar q$ pair plus the
nuclear ground state causes the $q\bar q$ component of the pion wave
function to dominate the reaction process. At  very  high beam
momenta, the pion breaks up into a $q\bar q$ pair with large
$\kappa_{t}$ well before hitting the nucleus. The dominance
of this starting point is verified in the present work. Note  also
the crucial feature that for values of $x$ which are not very small,
so that the leading twist approximation is valid for  the small dipole
- target interaction, a spatially  small wave packet of quarks and
gluons  remains  small as it moves through the target.
This leads to a dominance of the effects of large transverse momenta,
and allows the  factorization of the hard physics from the soft physics.
On the other hand, for very   small  values of $x$, the packet lives  so long
that it would expand to a normal hadronic size causing the initial state
interaction to  become similar to the soft one.
Moreover,  a rapid  increase of PQCD amplitudes with energy leads  to the
violation of QCD evolution equation at rather small values of  $x$,
and to the disappearance of the characteristic physics of  the
interaction of a small dipole. Thus QCD predicts different calculable
dependencies of the cross section of the 
diffractive dijet production on atomic number, on $\kappa_t$ and  $z$
in different regions of the $\ln \kappa_t^2/\Lambda_{QCD}^2$, $\ln 1/x$ plane.

Thus two large parameters  $s$ and ${\kappa_{t}}^2$ are present, and
this feature will enable us  to demonstrate the
dominance of   Feynman diagrams of a very  few specific topologies, and to
evaluate them.  The result of calculations can be represented in the form of
a generalized QCD factorization theorem,
valid for the set of Feynman diagrams corresponding to the leading
power of $s$ and the minimal power of ${1/\kappa_{t}}^2$  and $\alpha_s$
at fixed values of $\alpha_s\ln\kappa_t^2\Lambda_{QCD}^2$.
Our calculation relies heavily on the  well-known theoretical observation
that amplitudes of many high energy processes (such as that in
the parton model, multi-peripheral processes and  those involving the
Pomeron) are dominated by ladder diagrams \cite{Gribov,Feynman}. This
property has been proved using the approximation of including the terms of
lowest order $\alpha_s$ and all powers in  $\alpha_s \ln Q^2/\Lambda_{QCD}^2$
\cite{Gribov-Lipatov,Dok} and/or ~ $\alpha_s \ln 1/x$  \cite{Gribov,BFKL}.
The dominance of  ladder diagrams makes it possible to absorb the
effects of the leading terms  in $ \ln \kappa_{t}^ 2/\Lambda_{QCD}^2$
and $\ln 1/x$  into the  dipole-target interaction, and/or
into the target's skewed parton distributions. Furthermore, one may
classify and  analyze those diagrams  of leading power of $\alpha_s$
which are relevant for the pion transition into two jets.  This leads to
a selection rule: the  $t$-channel exchanges with  vacuum quantum numbers
(positive charge parity) should dominate. Thus we will calculate an
amplitude which is symmetric under the transposition $s\leftrightarrow u$.
Negative charge parity  contributions (such as effects of the odderon)
are neglected. There is another group of correction terms of the form
$\alpha_s \ln \kappa_{t}^ 2/\Lambda_{QCD}^2$
which arises from  radiative effects in the pion wave function.
The ladder structure of the dominant diagrams makes  it possible to
include these terms into QCD evolution of pion wave function and structure
functions of the target will not change the structure of  formulas
deduced by  considering skeleton diagrams calculated within the
approximation of keeping leading powers of $\alpha_s$; their  only
influence is  to introduce the effects of evolution in $\kappa_t^2$
into the relevant parton distributions and into high $k_t$ behavior of
the pion wave function. Indeed, our theoretical analysis heavily relies upon
specific properties of skewed parton distributions
and of minimal Fock component of pion wave function.
Note also that, for  small values of $x$, and large values of $Q^2$,
the skewed parton distribution of  a target nucleon or nucleus
is calculable in QCD using the appropriate  evolution equation
and  initial diagonal parton densities \cite{FFGS}.
To simplify the calculations and  especially the separation of the scales, we
use an axial light-cone gauge, which reduces to the
$A_{-}=0$  gauge, in the target rest frame. This gives  a
high $\kappa_t$ behaviour of fermion propagator and  hard gluon
exchange amplitude which have no infrared singularities. In this
gauge, unphysical degrees of freedom are removed from the light cone
pion wave function
at least within the leading log approximation.
Consequently, the separation of momentum scales can be easily made. On the
contrary, in the gauge used in \cite{BL} the separation of scales and
therefore derivation of the QCD evolution equation, although correct,
is complicated by the need  to account for the cancellation of the infrared
divergences. We want to stress that in the calculation of amplitudes of hard
diffractive processes in the gauge where fermion propagators are
infrared divergent one, should first remove infrared divergences and
only then  consider partons in the non-perturbative wave function to
be on-mass-shell.

These technical considerations lead to some simple results for
situations, such as ours, in which
the momentum transfer to the target nucleus is very small
(almost zero for forward scattering). In this case,  the dominant
source of high momentum must be the gluonic
interactions between the pion's quark and anti-quark.
This is also justified in the present work.
Because $\kappa_t$ is large, the quark  and anti-quark must
be at small separations--the virtual state of the pion is a
point-like-configuration\cite{fmsrev}. But the coherent interactions
of a color neutral  point-like configuration are  suppressed (at fixed
$x_{N}$, $\kappa_{t}^2\to \infty$), for the processes which involve
small transfers of momentum to the target, by the cancellation of
gluonic emission from the quark and anti-quark \cite{flow1,bb,fmsrev}
and /or from the  $q\bar q g $ state.-see the discussion in the section II.D.

Furthermore, the strength of the interaction with the target is
proportional to the square of the transverse separation distance between the
quark and anti-quark. Thus the interaction with the nucleus is very rare,
and the pion is most likely to interact with only one nucleon. The
result is that in this process the initial $\pi$ and
the final $q\bar q$ pair do not get absorbed by  the target, as would
typically occur in a low momentum transfer process. Thus initial and final
state interactions are suppressed and color transparency unambiguously
follows. As the values of $x$  are  decreased, the qualitative physics
changes  gradually. The  increase of the effective size of the small
color-dipole leads to the increase of the influence of  initial state
interactions, and  to a contribution of nuclear shadowing which enters
at leading twist. For even smaller values of $x$, the leading twist
approximation breaks down.

Our  treatment of the reaction process in terms of a separate wave
function and interaction pieces provides a new example of how
the QCD factorization theorem  works for high energy processes
involving  two large variables: $s,\kappa_t$ \cite{fmsrev}.
For this coherent process, the forward scattering amplitude is
almost proportional to the number of nucleons, $A$, and the cross
section varies as $A^2$. The forward angular distribution is
difficult to observe, so one integrates the angular distribution,
and the  $A^2$ variation becomes
$\approx {A^2\over R^2_A/3+B_N} \propto {A^{4/3}\over 1+0.45 A^{-2/3}}
\approx A^{1.37}$. Here
$B_N\approx 4.5 GeV^{-2}$ is the slope of the $t$ dependence of the
cross section of a hard diffractive process as determined by
data for electroproduction of vector mesons. This very rapid variation
represents a  prediction of a very strong enhancement which occurs via
the suppression of those interaction processes which usually reduce the
cross section. Our interest in this curious process has been renewed
recently by exciting experimental progress \cite{danny}.

Three key predictions of our paper \cite{fms93} are confirmed by the E791
data:
\begin{itemize}
\item The result  from the E-791 experiment
comparing Pt and C targets  that the coherent
cross section for small momentum transfer
to the nucleus  varies as $\sim A^{1.55\pm0.05}$,
is  close to our predictions, see Section V. This variation is
much stronger than seen in  soft diffraction
of pions by  nuclei $\sim A^{0.8}$\cite{fmsrev,explain},
which  is qualitatively different from the behavior $\sim A^{1/3}$
suggested in Ref.~\cite{bb}.
This  A-dependence is somewhat more rapid  than that predicted
by color transparency theory for $A\to \infty$. For moderately large
values of $A$, small effects
discussed in Sect.~V  tend to increase the $A$
dependence. This may be understood as a result of  the experimental
trigger  not excluding a   small but calculable
admixture of the effects of nuclear disintegration processes
which lead to similar dependence of cross section on $t$. Furthermore, it
is an unusual feature\cite{fms93} of the present process that
final state interactions of the point-like configuration  tend
to increase the $A$-dependence. Sect.~V  also contains
a  discussion of the  changes of the A-dependence due to the effects of
nuclear shadowing on the gluon density.

\item  The dependence of the cross section $\propto z^2(1-z)^2$
on the fraction of momentum $z$ carried by one of the jets
is consistent with our prediction.
\item
The cross section $d\sigma/d\kappa^2 $ falls as $\kappa_t^{-n}$
with $n=   10.2 \pm 0.4\;({\rm stat})\pm 0.3\;({\rm sys}),$
for $\kappa_t\geq 1.25$GeV  and as $n=7.5\pm 2.0$
for $\kappa_t\geq 1.8$ GeV. This  should be compared to the
prediction of $n= 8$  \cite{fms93}. For smaller values of  $\kappa_t$
soft  QCD phenomena, such as production  of  $q\bar q g$ jets,
should be important. See the discussion in  Sect.~\ref{3jets}.

\end{itemize}

The purpose of the present work is to rederive and confirm   our
earlier theoretical results with a more extensive analysis. The
derivation of our leading term\cite {fms93} directly from QCD by
generalizing a QCD factorization theorem of Ref.~\cite{cfs} was presented in
Ref.~\cite{hallertext}, and this is explained more fully now.  But
here we go further by verifying the assumption that the
point-like-configuration is indeed formed well before the projectile
reaches the nucleus. In the derivation we shall explain that several
different amplitudes, which seem to be of the same or lower order in
$\alpha_s$ as the leading one described above really are very small
after proper account of the suppression of radiation collinear to pion
momentum direction.

We also update our study of the leading multiple-scattering correction,
which is positive because the strength of the final state interaction
decreases with  decreasing  size of the dipole \cite{fms93},
and we study the most important competing
electromagnetic process. Some specific features of the experimental
extraction of the coherent part of the cross section are also explained.
Still another feature  involves the
soft interaction between the dipole and  the target.
This was at first derived  to be proportional to the gluon
density of the nucleus\cite{bbfs93}. However, there is a non-zero momentum
transfer to the nucleus, so it is actually the skewed gluon density that
enters.  The skewedness of gluon distribution in the nuclear target
leads to a small,  calculable correction to the predicted A dependence
\cite{guzey} and absolute value \cite{FFGS} which changes the detailed nature
of our results but not the qualitative features.

Our main results are summarized by the following formula, valid in
the leading log approximation, for the differential cross section of
diffractive dijet production by nuclei:
\begin{eqnarray}
& & {d\sigma(\pi + A\to 2jet +A)(q_t=0)\over dt dz d^2 \kappa_{t}}=
\frac{(1 +\eta^2)}{4\pi (2\pi)^3}
{\left(\int d^2r_t  {d\beta\over \beta}
\exp i(\bbox{\kappa_t\cdot r_t}) \cdot \right.}
\nonumber\\ & &
{\left.(\alpha_s \pi^2/3) \int d^2 k_{1t}
\left[2\chi_{\pi}(z,r_t) -
\exp(-i ({\bf r_{t}}\cdot{\bf k_{1t}})
\chi_{\pi}(z-\beta,r_t) -
\exp(i({\bf r_{t}}\cdot{\bf k_{1t}})
\chi_{\pi}(z+\beta,r_t) \right]
\frac{f_{A}(x_{1},x_{2},\beta s,k_{1t}^2)}
{k_1^2 k_2^2} \right)^2},
\label{unint}
\end{eqnarray}
where by definition ${\bf r_t}$ is the  transverse distance between
the pion's quark and anti-quark,
${\chi_{\pi}(z,k_t)\over d_f(k_t)}\equiv \int d^2 r_t
\exp{i(\kdotr)}\chi_{1,\pi}(z,r_t)$
and $x_1G_A(x_1,x_2,Q^2)=\int_0^{Q^2}d^2l\; \int_{\beta_0}^1
{d\beta\over \beta} d^2l {f_A(x_1,x_2,\beta s,l^2)k_{1t}^2\over k_1^2 k_2^2}$
($f_A$ can be denoted as the unintegrated skewed nuclear gluon
density), $k_i$ are four momenta of two exchanged gluons, (see Fig.~1),
and $k_{1t}=\kappa_t-l_t$.

 Here $d_f(\kappa_t^2)$ is
the renormalization factor for the  quark Green function $S_f(k)$ in the hard
regime where
\begin{eqnarray}
S_f(k)={d_f(\kappa^2)\over \hat{k}},
\label{greenfunction}
\end{eqnarray}
for $k^2=\kappa^2$.
The quantity  $\beta_0$ is a complicated  function
involving the transverse momenta of the quarks within the
pion, and in the region giving the dominant contribution
$\beta_0 \propto \kappa_t^2$/s.
The quantity $\eta$ is the ratio of the real to
imaginary part of the $q\bar q$-target scattering amplitude.
In Eq.~(\ref{unint}), $\chi(z,r_t)$ includes both the
non-perturbative $q\bar q$ component and its high momentum tail. This function
therefore involves distances significantly
smaller than  average hadronic inter-quark distances. The
actual distances involved in  the largest contributions
is one of subjects investigated here.

Equation ~(\ref{unint}) is derived  in two steps. First we demonstrate
the  dominance of the Feynman diagrams of  Fig.~1 and then evaluate
these diagrams. The factorization of the hard perturbative QCD part,
related to the pion wave function, and $q\bar q$ pair arising from
softer QCD, described by skewed parton distributions (the dominance of
the diagrams of  Fig.~1) is another form of the QCD factorization theorem
derived in \cite{cfs} for  diffractive vector meson production in
deep inelastic scattering DIS. The end point contribution - the
Feynman mechanism -($z\propto {\Lambda_{QCD}^2\over \kappa_t^2}$)
is suppressed as compared to the leading term by a set of factors: one
power of  ${1\over \kappa_t^2}$,
the square of the  Sudakov-type form factor, by a form factor $w_2$
(which  accounts for  the very small probability to find a pion with $q$ and
$\bar q$ at {\it average} distances without a  gluon field) and by
the overlap integral with final state. A detailed analysis of the end
point  contribution will be subject of a separate
publication. It follows from QCD factorization theorems that the
amplitude of hard processes can be represented as the convolution
of the non-perturbative pion wave function and hard amplitude $T$.
The virtualities of all particles in the $s,u$ cuts of the amplitude
$T$ are large. Virtualities of those seemingly on-mass-shell particles
are
$\gg \Lambda_{QCD}^2$ but $\ll \kappa_t^2$. This is the condition which
dictates the dominance of perturbative tail in the pion wave
function at $\kappa_t^2\to \infty$ but fixed $x=\kappa_t^2/\nu$.
The amplitudes having different topology are radiative corrections
involving extra powers of $\alpha_s$. To elucidate the  underlining
physics we shall prove the dominance of
the amplitude $T_1$  (see Fig.~1)
by analyzing different contributions of many diagrams. A more general
proof will be given elsewhere. In the leading
$\ln{Q^2\over \Lambda_{QCD}^2}$ and $\ln {1\over x}$ approximations
the dipole description can be used to simplify the  above equation  to
the form:
\begin{equation}
{d\sigma(\pi + A\to 2jet +A)(q_t=0)\over dt dz d^2 \kappa_{t}}=
\frac{(1 +\eta^2)}{16\pi (2\pi)^3}
{\left[\Delta\left({\chi_{\pi}(z,\kappa_t)\over d_f(\kappa_t^2)}\right)
 {\alpha_s \pi^2\over 3} x_1 G_A(x_1,x_2,Q^2) \right]^2}.
\label{dijet}
\end{equation}
where $\chi_{\pi}(z,\kappa_{t})\equiv{4\pi
C_{F}}{\alpha_{s}(\kappa_{t}^2)\over \kappa_{t}^2} \sqrt{3} f_{\pi}z(1-z)$,
$\Delta$ is the Laplacian in  $\kappa_t$ space, $z$ is the fraction
light-cone (+) momentum  carried by the quark in the final state,
$x_{1}G_{A}(x_{1},x_{2},\kappa_{t}^2)$ is the skewed gluon density of
the nucleus, $x_{1},x_{2}$ are the fractions of target momentum
carried by exchanged gluons 1 and 2, $x_1-x_2=M^2_{2jet}/s$, $x_2\leq x_1$,
and $\eta=Re F/Im F$  (with $F$ as the dipole-nucleon scattering amplitude).
Note that the resulting $\kappa_t^{-8}$ dependence is  a consequence
of a kind of dimensional counting, as explained in Sect.~II.

It is necessary to discuss the kinematic and dynamic limitations of
our analysis. We require high beam energies so that the point-like
configuration remains small as it passes through the nucleus, and we also
require that $\kappa_t$ be large enough so that the $q\bar q$ pair  actually
be in a point-like configuration. This situation corresponds to
$\kappa_t^2/s$ being held fixed for large values
of $\kappa_t^2$.  For the experiment of Ref.~\cite{danny}
$\kappa_t\approx 2$ GeV, and $s=1000\; {\rm GeV}^2$, so
$x_N  \equiv{2\kappa_t^2\over s}\approx .008$. There is another
kinematic limit in which $\kappa_t^2$ is fixed and
$s$ goes to $\infty$. At sufficiently small values
of $x_N$, less than about ${1\over 2m_N\;R_A},$
the situation is very  different because the $q\bar q$ dipole
system is  scattered by the collective gluon field of the nucleus.
Nuclear modifications (enhancement) of the nuclear  gluon density
actually occurs  at  larger values of $x_N$   corresponding to
$x_N\sim {1/( 2m_N\;r_{NN})} \sim 0.1$ (where $r_{NN}\sim 2$ fm
is the  mean inter-nucleon distance in nuclei\cite{FLS90,Pirner,Eskola}).
But for values such that  $x_N\leq {1\over 2m_N\;R_A}$
the  nuclear gluon field is expected to be  shadowed,
leading  to a gradual disappearance
of color transparency  (at a fixed scale ($\kappa_t^2$)).
This is the onset of perturbative color opacity \cite{fms93,fs99,NMC}.
At even smaller values of $x_N$ a new phenomenon has been predicted --
the  violation of the  QCD factorization theorem  \cite{fks}. Our
present analysis is not concerned with this region of extremely small $x_N$.
Another interesting phenomenon is the  possibility of probing
the decomposition of quark distribution amplitude
in terms of  Gegenbauer polynomials
at sufficiently large values of $\kappa_t^2$.

Some general features of our analysis appear in several different
Sections, so it is worthwhile to discuss these here.  The calculations
of several amplitudes are simplified by the use of a general theorem.
In the leading order in $1/\kappa_t^2$, the interaction of the $q\bar q$
occurs via  the exchange of a two-gluon ladder with the target. It is
important to note that the interaction of the
$q\bar q$ pair with the target via the exchange of a larger number of
gluons is suppressed by  powers of $1/\kappa_t^2$. The proof of this
statement follows \cite{cfs} and  heavily uses Ward identities.
(To visualize the  similarity with the situation considered in
Ref.~\cite{cfs}, it is instructive to neglect the effects of  the
odderon contribution, which is in any case small. Accounting  for
symmetry of amplitude on the transposition: $s\leftrightarrow u$
gives the possibility to consider the  amplitude for the process
$q\bar q +T\to \pi +T $, and to repeat the reasoning of \cite{cfs}
by parametrizing the momenta of exchanged gluons along
dijet total momentum.) The QCD factorization theorem \cite{cfs}
predicts also that the interaction with the target via $t$-channel exchange
by $q\bar q$ pair (which is expressed in terms of  the skewed  quark
distribution) is not small at $x\geq 0.1$
and moderate $\kappa_t$ where gluon distribution is not large
\cite{cfs,gv}. At smaller $x\leq 10^{-2}$ and
$\kappa_t^2\geq 1.5 GeV^2$  where the gluon distribution is large,
as a result of $x$ and $Q^2$ evolution,  this term is a small
correction to the exchange of the two-gluon ladder, see Eq.~(\ref{eq:1.27}).
So in this paper, we shall neglect  this term.
The two gluons are vector particles (bosons) in a color
singlet state, so the dominant two-gluon exchange amplitude
occurs in a channel which
has  positive charge and  spatial parity, and  is therefore
even under crossing symmetry. Given this even
amplitude, and the condition that we consider high energies
$\nu\equiv 2p_{\pi}m_N$ and  fixed small values of the momentum
transfer $t$ to the target, we may use the dispersion relation over
invariant energy $s$ at fixed $t$ and fixed momenta of two jets
to reconstruct full amplitude via the amplitude cut over
intermediate states in s and u channels. In difference from
the amplitudes of hard diffractive meson production, the
amplitude for dijet production may have an imaginary part which
varies with  $M^2_{2jet}$ also. At the same time, within
the approximations made here,  the  cut amplitude coincides with
the imaginary part of the full amplitude. Furthermore, the relation
discussed below makes it possible to reconstruct the real part of 
the amplitude
from the  imaginary part. For an amplitude corresponding to a slowly
growing total cross section
($A\propto s^{\alpha} , \alpha \sim 1$) the relation is
\cite{real}
\bea
{{\rm Re} A(\nu,t)\over {\rm Im} A(\nu,t)}
={\pi\over 2}{\partial \over \partial
\ln{\nu}}\ln{{\rm Im} A(\nu,t)\over \nu}
\label{real}.
\eea
This means that we may simply calculate the imaginary parts of any contribution
to the scattering amplitude, as a function of
$s$ and $u$, with the full amplitude
obtainable from Eq.(\ref{real}). Furthermore, the imaginary part
of the scattering amplitude, ${\rm Im}A$  varies
rather slowly with $\nu$, leading to a small value of
${\rm Re}A/{\rm Im}A$. Thus  ${\rm Im}A$ dominates in the sum of diagrams.
The possibility of considering only the imaginary part of the scattering
amplitude  simplifies the calculations enormously. The relevant
intermediate states are almost on the energy-shell (virtuality of
quark, anti-quark, gluon $\ll \kappa_t^2$) and one can use conservation
of four-momentum to relate the momentum of the relevant
intermediate states to that of the initial state.

There is another enormous simplification which is related   to the issue
of gauge invariance. The pion wave function  is not gauge invariant, but
the s,u cut parts of the amplitude $\pi +g\to JJ+g$, for two  gluons
in a color singlet state, are calculable in terms of amplitudes of
sub-processes where only one gluon is off mass shell. For such amplitudes
the  Ward identities\cite{t'hooft}-- the conservation of color current--
have  the same form as the conservation of electromagnetic current in QED. In
QED the current conservation identity has long been used to simplify
calculations of high energy reactions\cite{Gribov}. We will often use the
Ward identities\cite{t'hooft} to extend the QED method to treat
various contributions to our process.
To be able to separate soft and hard scales one need to account for
the cancellation of infrared divergences introduced by using the light-cone
gauge $A_+=0$ \cite{BL}. Instead of accounting for this cancellation as
in \cite{BL} we choose different gauges for the description of parton
distributions in a nucleon, and for the amplitude pion
fragmentation. This is legitimate within the region of applicability
of leading log approximations. Within the chosen gauge, the hard gluon
exchange amplitude and fermion propagator have no infrared divergence.
Note also that after demonstrating the factorization of the  hard QCD
amplitude from the  soft  QCD amplitude,
we may and will approximate the soft part of the pion wave function by a
system of free $q\bar q$ \cite{hallertext}. However, we stress that
this approximation is dangerous  for evaluating  the pieces of
amplitudes dominated by soft physics especially if propagators contain
infrared divergences. In that case, this approximation violates Ward
identities and the energy-momentum conservation law.

Additional common features arise from considering the relevant kinematics.
In all of the two-gluon
exchange diagrams  we consider, Figs.~1-12, the target nucleon
of momentum $p$ emits a gluon of momentum $k_1$ and absorbs one
of momentum $k_2$.  Conservation of four-momentum gives
\bea
k_1-k_2=p-p'=p_f - p_\pi,
\eea
in which $p'$ and $p_f$ are the final momenta of the target and the
dijet. Taking the dot product of the above with $p_\pi$, for the large
pion beam momentum relevant here, leads to the relation:
\bea
x_1-x_2={m_f^2-m_\pi^2\over 2 p_\pi\cdot p}=
{m_f^2-m_\pi^2\over \nu},
\label {diff1}\eea
where \bea
x_{1,2}\equiv {k_{1,2}^+\over p^+},\label{x1x2}\eea
and where $m_f$ is the mass of the final 2jet
system: $m_f=M_{2jet}$. Within the parton model approximation
\bea
x_2>0, \label{x2}
\eea
except the region of very small $x_2$-- denoted the  wee parton region.
In the parton model this condition follows from the requirement
that a parton knocked out of a nucleon should be kinematically
separated from the rest of the  target.  Otherwise the  amplitude should be
suppressed by a power of $\kappa_t^2$ \cite{Feynman}. This
suppression disappears at sufficiently high energies for which the
parton wave function of a target develops wee partons.

Another important consequence of the positivity of ``mass''$^2$
of partons in  intermediate states Eq.~(\ref{x2}) is
that the fraction $\beta$ of the  pion's (+)
momentum carried by exchanged gluons should satisfy the condition:
\bea
1>\beta>0, \label{x3}
\eea
for our kinematics.
The restriction  $\beta \propto \kappa_t^2/\nu$ can be justified
within the leading $\ln1/x$ approximation only.

The results (\ref{diff1},\ref{x2},\ref{x3}) are significant because
they will be  used in the evaluation of other diagrams. In particular
the condition (\ref{x2}) is not fulfilled for the diagrams in which the
transverse momenta of quarks within the  pion wave function are significantly
smaller than the observed transverse momenta of jets. This means that the
quarks in the pion must have very high transverse momentum to satisfy
Eq.~(\ref{x2}). In this case, the quarks  are closely separated and
we may consider the configuration to be  a point-like configuration.
Such a restriction  is operative in the kinematics  for which the
target wave function has no wee partons, i.e. for sufficiently large
$x_N$. On the contrary, for sufficiently small values of $x_N$, the  sign
of $x_2$ becomes unimportant because the amplitude
does not depend on the sign of wee parton momentum.

It is also worth emphasizing that the dominance of small size
configurations in the projectile pion, so important to our analysis,
is closely related to the renormalizable nature  of QCD. This
renormalizability implies, as extensively discussed
below, that the selection of large transverse
momentum final-state  jets leads to a selection of the large transverse
momenta of the quarks in the pion wave function, and also to some
increase of transverse momenta of the exchanged gluons.

One also needs to realize  that the emissions in the in-state and absorptions
in the out-state combine in calculating  the usual parton density
to produce the renormalized parton density. Thus  it is necessary to
guarantee suppression of gluon radiation collinear to the pion direction
in the initial,  intermediate, and final states. Otherwise an 
exclusive process  will be additionally suppressed by powers of the
Sudakov type form factor. This is a stringent condition which
suppresses the contribution of all other diagrams except that of Fig.~1
because for small values of $x_N$,  the  time  and
longitudinal distance intervals  ($\sim {1}/({2m_Nx_N})$)
are easily long enough to accommodate the radiation of a gluon.
If a pion is in a spatially small quark-gluon component,
collinear radiation is suppressed because color is highly localized
in the plane  transverse to the pion momentum. As a result
(similar to  the case of  meson production by  longitudinally
polarized photons) there is  no Sudakov form factor type
suppression for such processes \cite{Frankfurt:1991rk}.
Note also that according to the  QCD factorization theorem
a  pion in a small size configuration consists of $q\bar q$ pair
accompanied by a coherent, relatively soft gluon field which
follows the valence quarks without violation of coherence.
This gluon field is included in the skewed gluon distribution.

Consider now the impact of the above-mentioned condition for the
interactions of a pion in a large size $q\bar q$ configuration. The
$q$, $\bar q$ and gluons which start off far apart must end up with a
q$\bar q$ pair close together in a final state without collinearly
moving gluons. In this case, the $q$ and $\bar q$ must undergo a high
momentum transfer without emitting gluons collinear to the pion direction.
But  such processes are well known to be exponentially suppressed by
double logarithmic Sudakov-type form factors. Only in the case of a
compact $q\bar q$ pair, of  a transverse size  commensurate with  the
virtuality of a gluon bremsstrahlung, would the gluon radiation
be small. A related suppression, evaluated in  Sect.~III  is
the very small probability $w_2$ of  finding  a pion with $q$ and
$\bar q$ at {\it average} distances without a  gluon field, if the
probe has a resolution $\kappa_t^2\gg \Lambda_{QCD}^2$. Note that
under these conditions, in the typical parton configuration, gluons
are experimentally observed to carry about $\sim 1/2$ of the pion
momentum. Another example is pion
scattering by a high momentum gluon field of a target. In the
intermediate state there should be a strong collinear radiation
along the pion direction because  the  color charge strongly changes
its direction of motion, and there is no color charge nearby to
compensate for this emission. This is similar to the effect of
filling a gap in the case of color unconnected hard processes like
Higgs production via $gg \to H$ in hadron-hadron collisions \cite{higgs}.

In considering hard exclusive processes,  one needs to  address
the problem of the end point contributions -- the so called Feynman
mechanism. We find that due to the color neutrality of the pion and
the effect of target recoil,  the amplitude for this mechanism is suppressed
by a factor $\propto 1/\kappa^2$ as compared with that of the
perturbative QCD mechanism.
The contribution of the Feynman mechanism is also suppressed by powers
of the Sudakov-type form factor and by the form factor $w_2$.

In previous papers\cite{fms93,hallertext} we have emphasized that
the amplitude we computed in 1993 is calculable  using perturbative
QCD. However, there are five other types of contributions which occur
at the same order of $\alpha_s$. The previous term in which the
interaction with the target gluons follows the gluon-exchange in the
pion wave function has been denoted by $T_1$. However, the
two gluons from the nuclear target can also be annihilated by the exchanged
gluon (color current of the pion wave function). This group of
amplitudes is
denoted as $T_2$. Another term in which the  interaction with
the target gluons occur before the gluon-exchange in the  wave
function of dijet has been denoted by $T_3$. There are also terms, denoted
as $T_4$, in which the interaction with the target gluons spans the entire
time between interactions with target gluons. This term corresponds to the
interaction of the $q\bar q g$ configuration with target gluons. Still
another amplitude, $T_5$,  describes the interaction of $q\bar q$
dipole with a target in non-leading order in
$\alpha_s \ln \kappa_t^2/\Lambda_{QCD}^2$.

Here is an outline of the remainder of this paper. Sect.~II considers
the $\kappa_t^2$ dependence of the  Feynman diagrams and selects
Feynman diagrams having the minimal power of $1/\kappa_t^2$ at fixed
values of $\alpha_s \ln \kappa_t^2/\Lambda_{QCD}$ in the lowest order
in $\alpha_s$. We found that the processes where small size $q\bar q$
configuration is prepared before the scattering dominate diffractive dijet
production. Subtle features of the arguments are discussed in 
Sect.~III, which is concerned with  the role of selection of exclusive
processes in the suppressing of the contribution of hard processes related
to  inter-quark transverse distances $\gg 1/\kappa_t$.  All of the terms
$T_{2-5}$ are shown to be negligible in the sense that they are smaller
than $T_1$ by at least a power of $\alpha_{s}$ or by a factor of
${\Lambda^2\over \kappa_t^2}$, or by powers of Sudakov-type
form factors and/or of a  form factor $w_2$, which is related to
the probability of finding a normal-sized $q\bar q$ configuration of the pion.
At the end of this section we demonstrate that the
contribution to high $\kappa_{t}$ dijet production resulting from the
scattering of a  large size  $q\bar q$ dipole by a large transverse momentum
($\approx \kappa_{t}$) component of the target gluon field is suppressed
at least by two  powers of the Sudakov-type form factor and by the $w_2$
form factor. Together sections ~II and III form the proof  of the QCD
factorization theorem for the diffractive dijet production. Rather general
arguments for the small nature of the amplitudes $T_{2-5}$ provided in
the Sects.~II and III are valid for  the case of the photon projectile as well.
This is because the contribution into the forward scattering amplitude
of dijet production due to direct photon coupling 
 to light quarks is $\propto$ bare mass of quark
and therefore small \cite{brodsky,diehl}. In the case of a charm quark
photoproduction, the  dominant term is given by a charm component of
the
direct
photon wave function.

Sect.~IV is concerned with the evaluation of the dominant
amplitude $T_1$, which has the form of the QCD factorization
theorem in the leading order in $\alpha_{s}$ and all orders
in $\alpha_s\ln (\kappa_t^2 r_{\pi}^2)$. The analysis performed in
Sections ~II-IV shows that the $z$ dependence of
the leading (over powers of  $k_t^2$) term in the amplitude of diffractive
dijet production  is given basically by the factor $z(1-z)$.
The nuclear dependence of the amplitude,
including a reassessment of the multiple-scattering  correction
of \cite{fms93}, and nuclear shadowing effects is discussed in
Sect.~V. Experimental aspects, including the requirements for
observing color transparency and the extraction of the coherent
cross section, are discussed in Sect.~VI. There is an electromagnetic
background term, which becomes increasingly more important as
$\kappa_{t}$ increases, in which  the exchange of a photon with the
target  is responsible for  the diffractive dissociation
of the pion. This process, which occurs on the nuclear periphery and
is therefore automatically free of initial and final state interactions,
is shown in Sect.~VII to provide a correction of less than  a few percent
contribution to the cross section at values of  $\kappa_{t},s$
of the experiment \cite{danny} but this correction rapidly decreases
with an increase in the value of $\kappa_t$.
A discussion of the implications of
observing color transparency as well as a summary and assessment of
the present work is provided in the final Sect.~VIII.

\section{ Selection of dominant Feynman diagrams for $\pi N\to N JJ$
in the leading order of  $\alpha_s$ and $1/\kappa_t^2$ }

The kinematic constraints due to the energy-momentum conservation play
an  important role in the evaluation of amplitudes of diffractive
processes. Therefore, we begin by deducing the necessary kinematical
relations and  introducing the  light cone variables  we use.
Our interest is in the scattering at nearly forward angles. We denote
momentum of the  pion  as $p_{\pi}$, and that of the target as
$p$. The three momentum of the nucleus in the
final state is $p_{fz} \approx M^2_{2jet}/2E_{\pi}=m_N x$. The
first relation is expressed in terms of the variables of the nuclear
rest frame, and the second is in the variables of IMF of a nucleus
where $x=M^2_{2jet}/\nu$. The mass of  the two jet system  is given by
$M^2_{2jet}=\frac{m_q^2+\kappa_t^2}{z(1-z)}\approx \kappa_t^2/z(1-z)$.
We neglect  the mass of quark $m_q$ as compared with the  large jet
momentum $\kappa_t$. The quantity
$\nu =2(p_{\pi} p)/A$ is the invariant energy of collision.
Our notation is that $z$ represents the fraction of the total
longitudinal momentum of the beam  pion carried by the quark in the
final state, and $1-z$ the fraction carried by the anti-quark.
The transverse momenta are given by $\vec{\kappa}_t$ and $-\vec{\kappa}_t$.
A is the number of nucleons in the nuclear target.
Our interest is in the kinematics for which the final state
nucleus remains intact. This means that minimal momentum
transferred to nuclear target $-t_{min}\approx p_{f,z}^2$
should be small: $-t_{min}R_A^2/3\ll 1$ i.e.
$x\ll {\sqrt 3\over m_N R_A} \approx A^{-1/3}/3$.
Here $R_A=1.1 A^{1/3}$ Fm  is the  nuclear radius.
For small values of $-t_{min}R_A^2/3$, the effect of
$t_{min}$ can  be easily accounted for because any form factor
of the target can be approximated as $\exp{\;tR_A^2}/6$.

For large enough values of $\kappa_t$, the result of the calculations can
be represented in a form in which only the $q\bar q$ components of the initial
pion and final state  wave functions are relevant in Eq.~(\ref{matel}).
See  also the discussion at  the beginning of section III.
This is because we are considering a coherent
nuclear process which leads to a final state consisting of a quark and
anti-quark moving at high relative transverse momentum.
It is necessary to examine the various momentum scales that appear in this
problem. The dominant non-perturbative component of the pion
wave function carries relative momenta (conjugate to the transverse
separation between the $q$ and $\bar q$) of the order of
$p_t\sim\frac {\pi/2} {\sqrt {2/3} \;(2r_\pi)}\approx 300 $ MeV.
This is much, much  smaller than the final state  transverse relative
momenta, which must be greater than about 2 GeV, the
minimal value required  to  experimentally define a jet.
The immediate implication is that the non-perturbative pion wave function,
which is approximately a Gaussian, cannot supply the necessary high relative
momenta. These momenta can only arise from the exchange of a hard gluon, and
this can be treated using perturbative QCD.

Restricting ourselves to Feynman diagrams having the leading power of
$s$, $\kappa_t^2$, $\alpha_s$ (at fixed
$\alpha_s \ln \kappa_t^2/\Lambda_{QCD}^2$ and/or $\alpha_s \ln1/x$)
and using a normal  non-perturbative wave function which rapidly
decreases with increase of the constituent transverse momentum gives the
possibility of  regrouping  the diagrams into blocks having a rather
direct physical meaning:
\begin{eqnarray}
{\cal M}(N)&=&
(T_1+T_2+T_3+T_4+T_5)\;.\label{mdef}
\label{tdef1}
\end{eqnarray}
Here the dominant terms $T_1$ and $T_{1b}$ of  Figs.~1 and 2 represent
a type  of impulse approximation, and we  shall examine  them first.
Next we evaluate the
possible role of color flow - of the interaction of target gluons with
with gluon exchanged between quarks in the pion wave function
-amplitude $T_2$.This amplitude is also expressed through the same pion
wave function as $T_1$.
The term $T_3$ corresponds to the final state interaction between
jets.  This  will be followed by the discussion of the physical
meaning of the other terms $T_4,T_5$ and the explanation of their smallness.
At this point and below  we rely
heavily on the fact that, if $\alpha_s\ll 1$ in the leading
$\alpha_{s}\ln\kappa_t^2/\Lambda_{QCD}^2$ and/or $\alpha_s\ln1/x$
approximation, the sum of dominant diagrams (but not each particular
diagram) has a ladder structure. In particular, various crossed
diagrams which have a different form are needed to guarantee local
gauge invariance for  processes with large rapidity gaps, and to ensure the
ladder structure of the sum of dominant diagrams. To derive
such a ladder structure of the sum, it is important to explore
crossing symmetry and the positive charge parity which follows from the
dominance of the t-channel exchange two-gluon state of vacuum quantum
numbers.  Accurate exploring of gauge invariance is necessary for
proving the ladder structure of the sum of Feynman diagrams. This
structure makes it possible to include terms varying as
$\ln 1/x$ and $\ln \kappa_{t}^2/\Lambda_{QCD}^2$,  related to
conventional QCD evolution, in skewed parton  distributions or in
skewed unintegrated gluon densities. In addition,  we need to calculate
the evolution of the pion wave function with transverse momentum
and the interaction of gluons with this wave function. So we shall
first  classify and calculate skeleton diagrams, and then  account
for the QCD evolution.

\subsection{Dimensional estimate of the initial state hard interaction}

In our previous papers we investigated the term $T_1$ of Fig.~1.
QCD is a non-Abelian gauge theory in which an exchanged gluon may
probe the flow of color within the pion wave function. So in the leading
order of $\alpha_s$, the requirements of gauge invariance
mandates that  we should consider also the related
set of diagrams where an exchanged gluon is attached to the
exchanged gluon in the $q\bar q $ component of the pion wave function, see Fig 2.
\begin{figure}[t]
\begin{center}
\epsfig{file=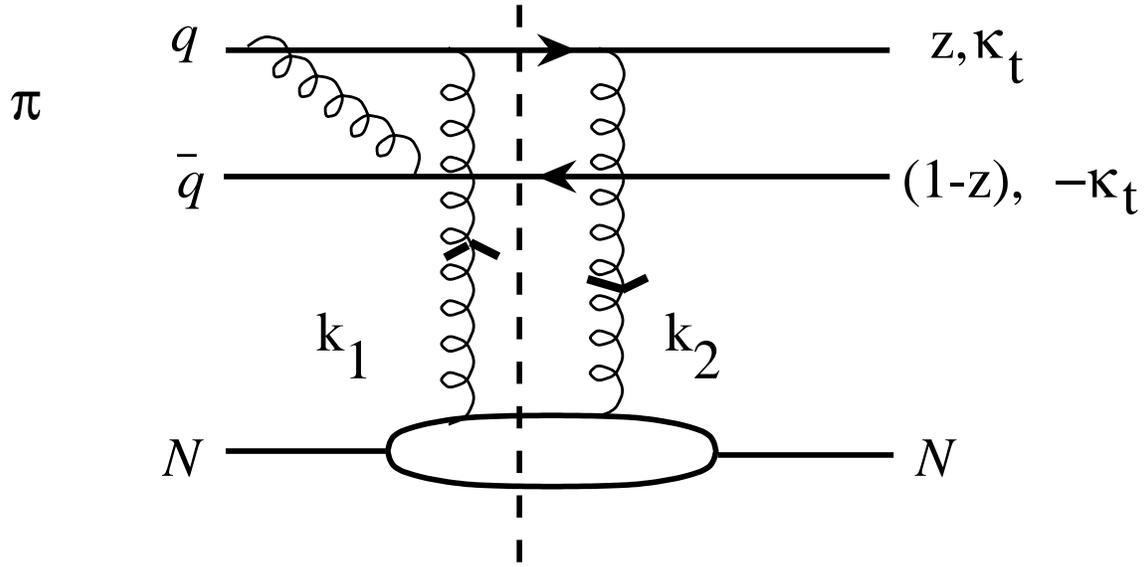, height=3.0in}
\vspace{0.3in}
\caption{ A contribution to $T_{1a}$. The high
momentum component of the pion interacts with the two-gluon
field of the target. The displayed diagram occurs along with its
version in which the gluons are crossed. Furthermore, there
are four diagrams for each term because each of the gluons can be
absorbed or emitted by either the quark or anti-quark of the beam pion.
Thus only a single diagram of the eight that contribute is   shown.}
\end{center}   \end{figure}
\begin{figure}[t]
\begin{center} \epsfig{file=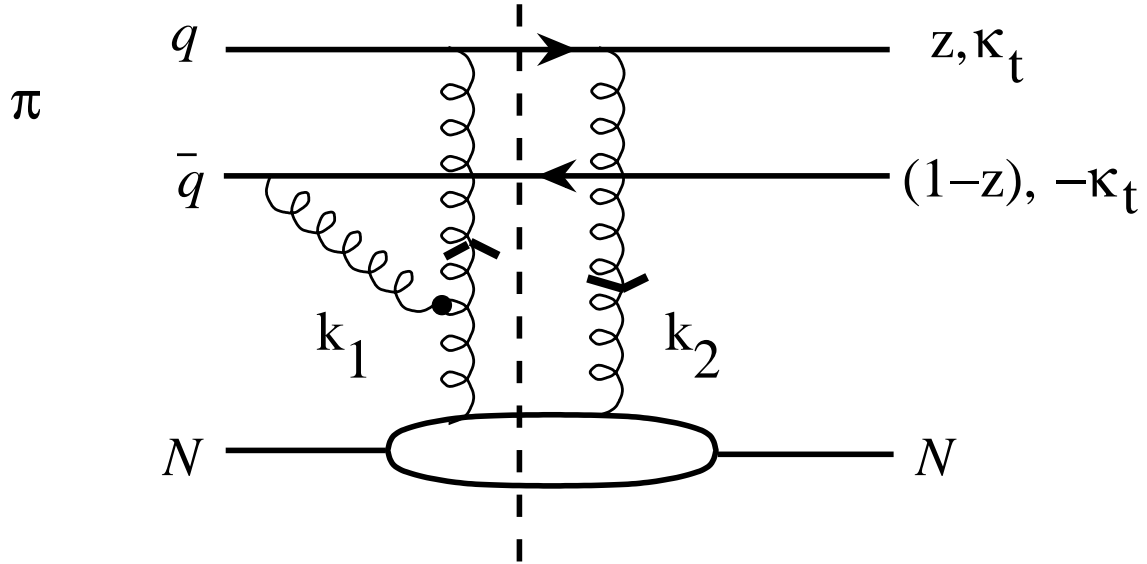, height=3.0in}
\vspace{0.3in}
\caption{ A contribution to $T_{1b}$. The high momentum gluon current
in the the pion wave function interacts with the two-gluon field of the
target. The displayed diagram occurs along with its version in which the
gluons are crossed, for different attachments of gluons in the pion wave
function. Thus only a single diagram of the eight that contributes is
shown. }
\end{center}   \end{figure}

In the evaluation of the terms of Figs.~1 and 2, with color flow we
use Gribov's observation
\cite{Gribov} that,  within the leading $\alpha_s \ln1/x$ approximation,
the polarization of gluons exchanged in the ladder is $\propto p$ where $p$
is the four momentum of the target. It is not difficult to show that the
same gluon polarization dominates in the calculation of QCD evolution
in the leading order in   $\ln\kappa_t^2/\Lambda_{QCD}^2$.
It is also    convenient to use the fact  that only one gluon
is off its  mass shell (in the leading logarithmic  approximation
considered in the paper). This  causes the equation that describes the 
conservation of color current to have  the same form as in QED.
We show here  that the use of QED-type Ward
identities\cite{t'hooft} (allowed in computing   the $s,u$ cuts
of the diagram, as explained in the Introduction) leads to the result
that only the transverse components of the gluon momenta $k_1,k_2$
enter in the final result for the  amplitudes $T_{1a},T_{1b}$.
Momentum factors, related to the contribution of
the  target
gluons, are included by definition in the skewed gluon distribution.

We examine the  part of  $T_{1a},T_{1b}$ that arises from
the exchange of the gluon $k_1$ and that gives  an on-shell
$q\bar q$ intermediate state. The result is
 \bea
T_1=T_{1a}+T_{1b} \propto A^\pi_{\mu \nu} d^{\mu\nu} d^{\tilde{\mu}\tilde{\nu}}
A^N_{\tilde{\mu}\tilde{\nu}}\cdots,
\eea
where $A^{\pi,N}$ represents the gluon emission amplitude of the
pion, nucleon and $d^{\mu\nu}$ arising  from the  propagator of
the gluons emitted or absorbed by the target. At high energies,
the gluon propagator can be represented as \cite{Gribov}
\bea
 d^{\mu\nu}\propto {2p^\mu p_\pi^\nu\over 2p_\pi\cdot p}\;,\eea
where $p^\mu$ is the nucleon momentum, so that
 \bea
T_{1a}+T_{1b}
\propto {2A^{\pi}_{\mu \lambda}p_{\mu}p_{\lambda}\over 2p_{\pi}\cdot p}
{2A^{N}_{\bar{\mu} \bar{\lambda}}p^{\pi}_{\bar{\mu}} p^{\pi}_{\bar{\lambda}}
\over 2p_{\pi}\cdot p}.
\label{t3a}\eea
We will denote  this presentation of the contribution of gluon
exchanges-as the Gribov representation because he was the first to
understand the dominance in the high energy processes of gluon
polarization $\propto p$, see \cite{Gribov}. Now we use current conservation,
\bea
A_{\mu \lambda}^{\pi}\cdot k_{1\mu}=0,\quad
\label{cc0}
  \eea
and employ Sudakov variables to describe the momentum $k_1$:
  \bea
  k^\mu_1=x_1\;p^\mu+\beta\;p_\pi^\nu+k_{t}.
\eea
We can determine the quantities $\alpha, \beta$ by
taking the dot product of the above equation with
either $p$ or $p_\pi$ and neglecting the relatively
small factors of the square of the pion or nucleon mass. This gives

\bea
x_1={k_1\cdot p_{\pi}\over p\cdot p_\pi},\nonumber\\
\beta ={\kappa_1\cdot p\over p\cdot p_\pi}.
\eea

Using these results in the current conservation
relation (\ref{cc0}) leads to the relation:

\bea
A^\pi_{\mu \lambda}\cdot p_{\mu}=-{\beta\over x_1} A^\pi_{\mu \lambda}
\cdot p_{\pi~\mu}-
{A^\pi_{\mu \lambda}\cdot \kappa_{1t\mu }\over x_1}.
\label{almost}\eea

The first and third terms of Eq.~(\ref{almost}), in
difference from the second, are proportional to $s$ and therefore
dominate over the second \cite{Gribov}. Thus we find
\bea
A^\pi_{\mu \lambda}\cdot p_{\mu}\approx -
{A^\pi_{\mu \lambda}\cdot \kappa_{1t\mu }\over x_1},
\label{yes}\eea
so that the exchange of the gluon $k_1$ gives an amplitude (\ref{t3a})
proportional to the small transverse momentum $\kappa_{1t}$. The net result
of these considerations is that the contribution of the gauge invariant set
of the diagrams including those of Figs 1, 2 takes the form:
\bea
T_{1a}+T_{1b}
\propto\;{2A^\pi_{\mu \lambda}\cdot k_{1\mu t} p_{\lambda}\over
(2k_1\cdot p_{\pi})}
\;\cdots\; {2A^N_{\mu'\lambda' }\cdot p_{\mu'}^{\pi}p_{\lambda'}^{\pi}
\over (2p_{\pi}\cdot  p)}\label{t3r}
.\eea
The same trick can be made with the second exchanged gluon:
\bea
T_{1a}+T_{1b}
\propto\;{4A^\pi_{\mu \lambda}\cdot k_{1~\mu t} k_{2~\lambda t}
\over (2k_1\cdot p_{\pi})(2k_2\cdot p_{\pi}) }
\;\cdots\; A^N_{\mu'\lambda' }\cdot p_{\mu'}^{\pi}p_{\lambda'}^{\pi}
\label{t4r}
.\eea

The factors involving $k_{it}$ will be absorbed into the definition
of the skewed gluon distribution of the nucleon. The useful result is
in the denominator of Eqs.~(\ref{t3r},\ref{t4r}), because
\bea
2k_i\cdot p = x_i\; s.
\eea
So
\bea
T_{1a}+T_{1b}
\propto\;{4A^\pi_{\mu \lambda}\cdot k_{1\mu t} k_{2\lambda t}
\over (x_1\nu)(x_2\nu) }
\;\cdots\; A^N_{\mu'\lambda' }\cdot p_{\mu'}^{\pi}p_{\lambda'}^{\pi}
\label{t5r}
.\eea
This formula in which the  amplitude is expressed in terms of exchanges by
transversely polarized gluons is the adjustment to QCD of the
Weizsacker-Williams (WW) method of equivalent photons\cite{WW}, of 
the Gribov
derivation of reggeon calculus \cite{Gribov}, and of the Cheng \& Wu's
impact parameter representation\cite{TTWu}. So below we will denote
such a formula as the Weizsacker-Williams representation of equivalent gluons.
For the calculation of the dominant amplitude $T_1$, such a trick is not
very useful because effectively $x_2\ll x_1$. So in this case
we shall use the Gribov representation for the second exchanged gluon.

Let us first perform power counting for the sum of terms
$T_{1a}+T_{1b}$ within the Gribov representation for the amplitude.
The number of strongly virtual propagators in  Fig 1 is two and the number of
large transverse momenta in the numerator from the vertexes is two. An
additional factor $1/\kappa_t^2$  follows from the cancellation of the
sum of leading diagrams because of the color neutrality of the
pion. In the following we include the factor $\alpha_s /\kappa_t^2$
as part of the high momentum component of pion wave function. Slowly
changing factors such as
$\alpha_s, \ln x, \ln (\kappa^2/\Lambda_{QCD}^2)$
are included by definition in the skewed gluon distribution $xG_A$
and into the pion wave function. Finally we obtain:
$T_{1a}\propto \alpha_s \chi_{\pi}(z,\kappa_{t}^{2})
x_1G_A(x_1,x_2,\kappa_{t}^{2})/\kappa_{t}^2$.
This result can be easily proved within the WW representation also. Note also
that the comparison of the above result with the WW representation
shows that the kinematical region around $x_2\approx 0$ gives a  negligible
contribution to the integral over $k_{it},\beta,x_1$.

The contribution of the color exchange current - the term  $T_{1b}$- is
suppressed in the light-cone gauge as compared to $T_{1a}$ by  a power
of $\kappa_t$. To  estimate the  $k_t$ dependence of the diagrams, we
shall use the WW representation of the  gluon exchange  between quarks
in the initial pion and the target and the  Gribov representation for gluon
exchange with the target in the final state. Within this
representation  for the sum of diagrams, the cancellation between
diagrams  leading to the number of powers of ${1\over \kappa_t^2}$ is
accounted for in a straightforward way. So the contribution of energy
denominators gives: ${1\over (\kappa_t^2)^3}$. Additional
${1\over x_1\nu} \propto {1\over k_t^2}$ as compared to the product of
gluon propagators is a result of cancellations between diagrams due to
color neutrality of pion and final state of two jets. Account of gluon
momenta in the nominator gives 0  when both  gluons in the gluon color
current in the pion wave function are  longitudinally polarized:
$(k_{ 1t})_r p_{\mu,\pi} p_{\lambda,\pi} g_{\mu,\lambda}^r =0$.
The contribution in the denominator when  one of the gluons  in the gluon
color current in the pion wave function is longitudinally polarized,
while  the second  is transversely polarized, also vanishes. This is
because forward scattering cannot change helicity. The proof of this
statement follows from the combination of  the light cone gauge
condition: $A_{-}=0$ and the fact that the leading power of $s$ is given by
the $+$ vertex for the interaction of aa nuclear gluon with a quark.
Accounting for the color neutrality of the pion wave function and of the wave
function of the dijet final state, as well as the 
anti-symmetry of 3 gluon vertex
is also important.

The use of identity:
$(k_{1t})_r \kappa_{\mu,t}   \kappa_{\lambda,t} g_{\mu,\lambda}^r =0$
helps to prove that the leading contribution into $T_{1b}$ is 0 when
both gluons in the gluon color current in the pion wave function are
transversely polarized. Here $k_{1t}$ is the transverse momentum of
a gluon exchanged with target and $\kappa_t$ is 
the transverse momentum of the jet. 
$ g_{\mu,\lambda}^r$ is the Yang-Mills three-gluon vertex.

The physical meaning of the obtained result for $T_{1a}+T_{1b}$
is that the interaction of quarks in the pion wave function with the
target gluon with a relatively low virtuality is dominated by
distances significantly larger than that involved in the pion wave function,
and  by the target gluon interaction  with external lines of the
amplitude without a gluon. This is the generalization to QCD of a
theorem proved for  QED by Low \cite{flow}.

For completeness we present the result of the calculation of Feynman diagrams
for $T_1$ in the leading
$\alpha_s \ln {\kappa_t^2\over \Lambda_{QCD}^2}$ approximation  where
the integration over fraction of pion momentum $\beta$ transfered to
the target gluon by quarks in the pion wave function is not performed. The
derivation is rather close to that in  \cite{bfgms} because in the
leading $\alpha_s \ln {\kappa_t^2\over \Lambda_{QCD}^2}$ approximation
the same polarization of target gluons dominates as in the leading
$\alpha_s \ln 1/x$ approximation. The deduced
formula is actually very similar to the formula deduced in \cite{bfgms}
for  hard diffractive vector meson production where in the leading
$\ln1/x$ approximation the integral over the unintegrated gluon density is
replaced in the final step by the gluon density.
So we will not repeat the detailed evaluation made in \cite{bfgms}.

The result is:
\bea
&&T_1/\nu=\int \alpha_s (2(2\chi(z,k_t^2)- \chi(z+\beta,k_t^2)-
\chi(z-\beta,k_t^2)) + \Delta (\chi(z+\beta,k_t^2)+
\chi(z-\beta,k_t^2)))\cdot 
\nonumber \\
&& \cdot {1\over d_f(\kappa_t^2)}
{k_{1t}^2\over 2} F^2{(3)}\pi^2
{d\beta\over  \beta} d^2 k_{1t} \frac{f_T} {k_1^2 k_2^2}.
\label{T_1}
\eea
Here $\Delta$ is the two dimensional $k_t$ space Laplacian operator
which acts on the pion wave function,
$f_T=\frac{A^T_{\mu,\lambda}k_{1t,\mu}k_{1t,\lambda}}{\beta\nu (2\pi)^4}$.
The first term of Eq.~(\ref{T_1}) is small and we  shall explain why
we neglect it.
The factor: ${1\over d_f(\kappa_t^2)}$ follows from the definition of
the pion wave function and from the definition of hard amplitude in terms of
series over the powers of the running coupling constant.
The  amplitude $T_1$ can be simplified by using the
leading $\ln1/x$ approximation. In particular,  one may express the
amplitude in terms of the gluon distribution \cite{bfgms}. This gluon
distribution is however different from the usual gluon distribution
determined from DIS processes because of the significant difference
between the masses of the initial pion  and final two jet systems. The
necessary skewed gluon distribution is calculable for small values of
$x$ as the solution of QCD evolution equation, using the  ordinary
diagonal gluon distribution as the initial condition \cite{FFGS}.
Here, in a fashion similar to \cite{bfgms}, we approximate:
$xG=\int d^2 k_{1t} {k_{1t}^2\over k_1^2 k_2^2} \int^1_{\beta_0}
{d\beta\over \beta} f_T$, so that $f_T$   can be denoted the
unintegrated gluon density. Equation~(\ref{T_1}) is a version of
Eq.~(\ref{unint}), in which the dipole approximation (keeping terms of
the order  $r_t^2$ in the bracket
$\left[2-\exp(-i ({\bf r_{t}}\cdot{\bf k_{1t}}))-\exp (i({\bf r_{t}}
\cdot{\bf k_{1t}}))\right] $
in Eq.~(\ref{unint})) is used. This  approximation is reasonable even
if $(k_{it}r_t)$ is comparable or even larger than 1 (cf. discussion
in \cite{bfgms} after Eq.~(2.20)).
Another useful approximation involves the value of $\beta$ appearing
in the argument of the pion wave function.  The upper limit of
integration over $\beta$ is dictated by  energy conservation  to be
$\beta\leq 1$.   For $z \sim 1$ one gets a further restriction
that $\beta \leq 1-z$. But in the leading
$\log {\kappa_t^2\over \Lambda_{QCD}^2}$
approximation, the condition for the region of integration in $\beta $ is
more severe. It is given by the requirement that the contribution of
the 
target gluon longitudinal momentum into its propagator
$\approx \beta M^2_{2jet}$ should satisfy conditions:
$\beta M^2_{2jet} \approx {k_{1t}^2 \ll M^2_{2jet}}$.
Hence in Eq.(\ref{T_1}) it is legitimate to neglect  $\beta $ in the
argument of the pion  wave function,
and to keep in Eq.(\ref{T_1}) the term $\propto \Delta$ only.
The value of the  lower limit of
the integration: $\beta\geq \beta_0$ is obtained from  the energy
conservation law, and from the QCD evolution which  effectively
suppresses the contribution of the  region $k_1^2/\beta_0 \nu \sim 1$.

Elastic interactions between high $\kappa_t$ $q$ and $\bar q$
in the final state may lead to an
infrared contribution which is exactly canceled in the probability summed
over the final state gluon radiation. It is important that the complete
nature of final states allows the term $T_1$ to  account for the
final state radiation and the space-time evolution of the $q\bar q$
pair in the final state. This phenomenon is a familiar feature
of the theoretical analysis of  the fragmentation of the small size
wave packet, of  $e+\bar e\to hadrons$.

\subsection{Meson Color Flow Term--$T_2$}

The $T_2$ or  meson-color-flow term, of Fig.~3 arises  from the $q\bar q g$
intermediate state, or from  the attachment of both target gluons to the
exchanged gluon or to the exchanged gluon and the quark(anti-quark).
\begin{figure}\begin{center}
\noindent
\epsfig{file=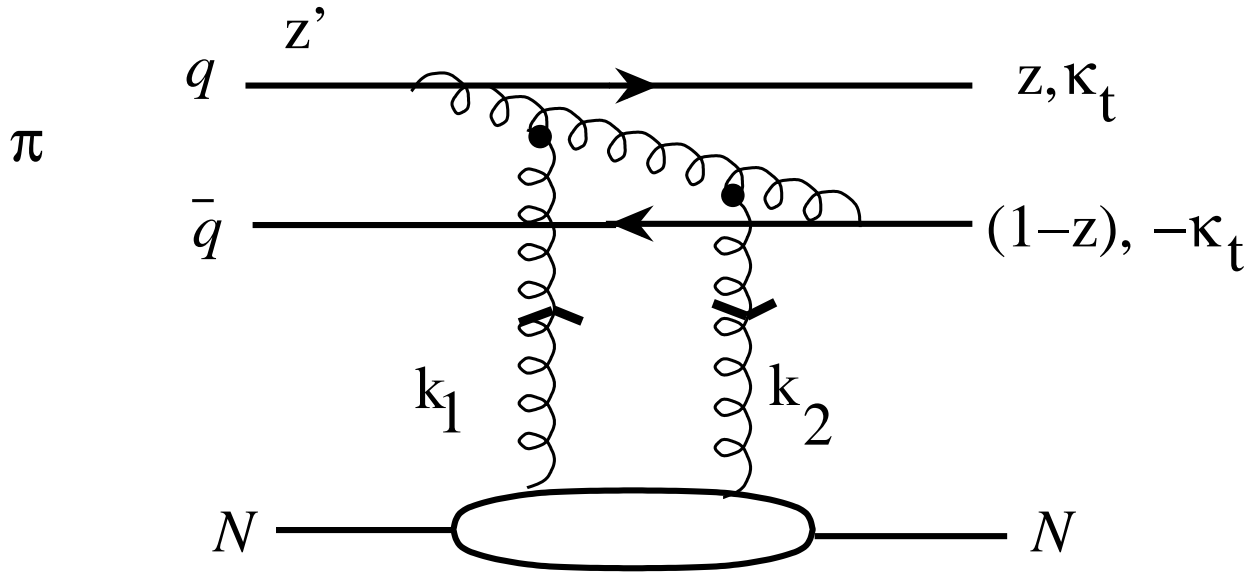, height=3.0in}
\vspace{0.3in}
\caption{
Contribution to  $T_{2a}$ of $q\bar q g$ intermediate state. The exchanged
gluon interacts with  with each of the target gluons. There is also a
diagram in which the gluons from the target are crossed, and another two
in which the exchanged gluon is emitted by the anti-quark. Only a single
diagram of the four that contribute is  shown.}
 \end{center}   \end{figure}

The diagrams of Fig.~3, considered in this section, have the same topology as
the diagrams included in the term $T_1$ and therefore have the same power of
${\Lambda_{QCD}^2\over \kappa_t^2}$, and  $\ln 1/x$. In particular,
they are included in $T_1$  if
$l_t^2 \ll \kappa_t^2$.
(To visualize the relationship between $T_1$ and $T_2$ it is useful to move
down the point of attachment of target gluons to the gluon in the pion wave
function in the diagram Fig.~3.)
In this case, a significant contribution may arise only from the perturbative
high momentum tail of the pion wave function. The effects of the
gluon-gluon interaction may then be included as part of the target  gluon
distribution. In the case when the square of the transverse momenta of the
gluon in the pion wave function is $\propto \kappa_t^2$, this diagram
can be considered as a non-leading order correction (to  the
dipole-target interaction  cross section) in $\alpha_s$ at fixed
values of $\alpha_s \ln {\kappa_t^2\over \Lambda_{QCD}^2}$. In contrast with
the term  $T_2$, the definition of $T_1$ contains a requirement that
the  transverse momenta of gluons attached to quark lines should be
much less than $\kappa_t^2$, and that the pion wave function includes
its perturbative tail. For the terms of Fig.~3, the non-perturbative
pion wave function  cuts off large quark momenta in the pion
wave function. Thus gluonic transverse momenta  are $\approx
\kappa_t^2$,   but transverse momenta of target gluons are still small:
$\kappa_{it}^2\ll \kappa_t^2$.  The contributions of other diagrams
with an almost on shell $q\bar q g$ intermediate state, see  Fig.~4,
are  not suppressed by a power of ${\Lambda_{QCD}^2\over \kappa_t^2}$.
\begin{figure}\begin{center}
\noindent
\epsfig{file=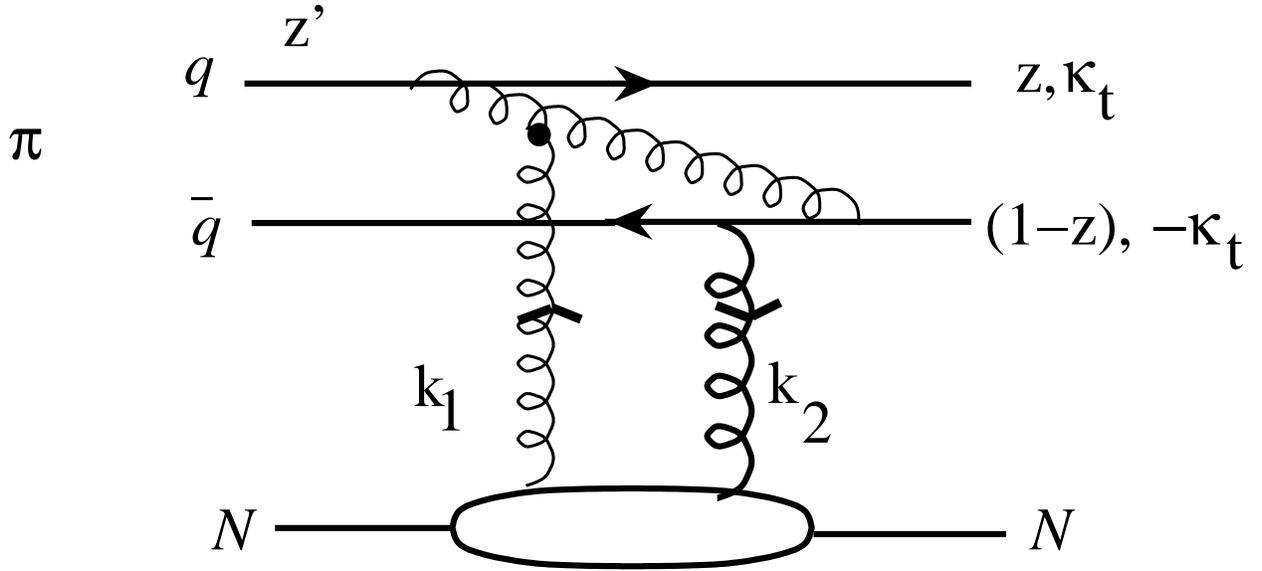, height=3.0in}
\vspace{0.3in}
\caption{
Contribution to $T_{2b}$ from $q\bar q g$ intermediate state. The interaction
of one target gluon field with exchanged gluon in the intermediate
states. There is also a diagram in which the gluons from the target
are crossed, and another group in which the exchanged gluon is emitted by
the anti-quark. Only one of 16 diagrams that contribute is shown.}
 \end{center}   \end{figure}

It is easy to calculate the sum of the terms $T_{2a,2b}$ in the
framework of  the
Gribov representation where dominant contribution is given by the 
polarization of a 
target gluon which is $\propto$ target four momentum. Most
straightforward calculation is for the ratio ${T_2\over T_1}$
because in this ratio all factors except the $z$ dependence
are canceled out. Really
quark-gluon vertexes when gluons are attached to quark(anti-quark)
lines in the pion wave function are the same for the terms $T_1,T_2$.
In the light cone gauge all factors from the vertexes for the
interaction of target gluons with a gluon  in the initial and final
state wave functions are effectively the same as for a gluon
interaction with quarks  except for the  Casimir operator of the
color group in the octet and the triplet representation. A subtle point
of calculation is to evaluate the $z$ dependence of this
ratio. For certainty in the evaluation of term $T_2$ we assume that
the 
nonperturbative pion wave function is equal to the asymptotic one.

Thus the ratio is determined by the color content of color
flow in the pion wave function and the quark color
and by the dependence of energy denominators on the fraction of pion
momentum carried by quarks and gluons. So
\bea
{T_2\over  T_1}={F^2(8)\over F^2(3)}
\left(
-1+ {1\over z(1-z)}+{z\over (1-z)^2}\ln z+
{(1-z)\over z^2}\ln(1-z))\right)
\label{t4rat}
\eea
Here $F^2(i)$ (for $i=8,3$ is the Casimir operator for octet and triplet
representations of color group $SU(3)_c$. The ratio $T_2\over T_1$
is $\approx 0.5$ for $z=1/2$, remains nearly constant for $|z-.5|\leq
0.3$ and increases to 9/8 at z=0,1.
This term is additionally suppressed by the  Sudakov type form factor and by
the form
factor $w_2$ -see the discussion below.

\subsection{Final state interaction of $q\bar q$ Pair--$T_3$}

The interaction with the target gluons may occur before the interaction
between quarks in the final state, and the related amplitudes are  denoted
as $T_3$, see Figs.~5 and 6.
\begin{figure}[t]
\begin{center}
\epsfig{file=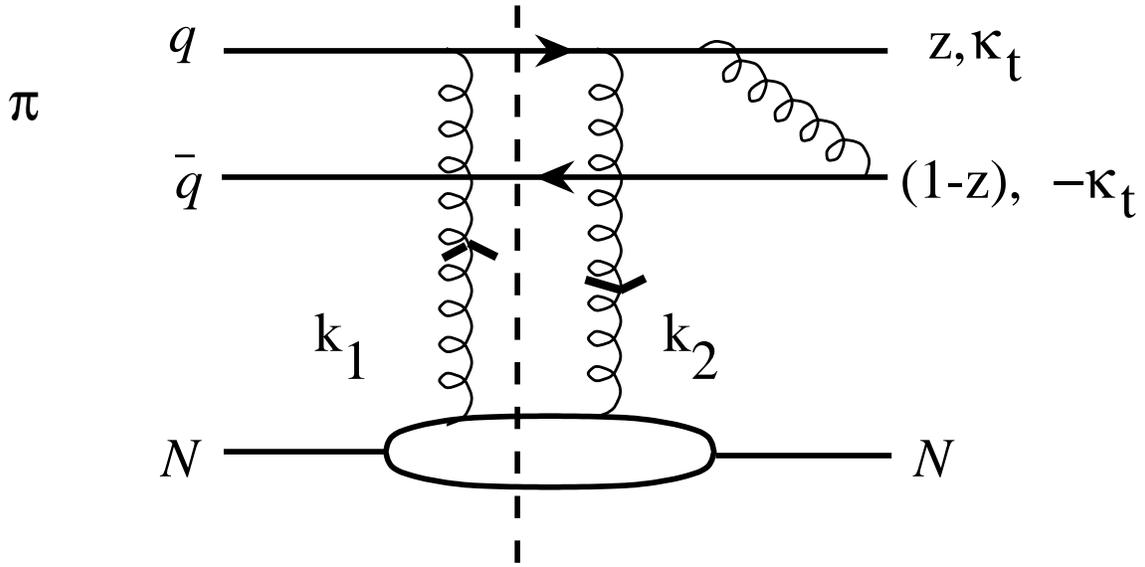, height=3.0in}
\vspace{0.3in}
\caption { Contribution to $T_{3a}$. The high momentum component of the final
$q\bar q$ pair  interacts with the two-gluon field of the target. Only a
single diagram of the  eight that contribute is  shown.}
  \end{center}  \end{figure}

\begin{figure}[t]
\begin{center}
\epsfig{file=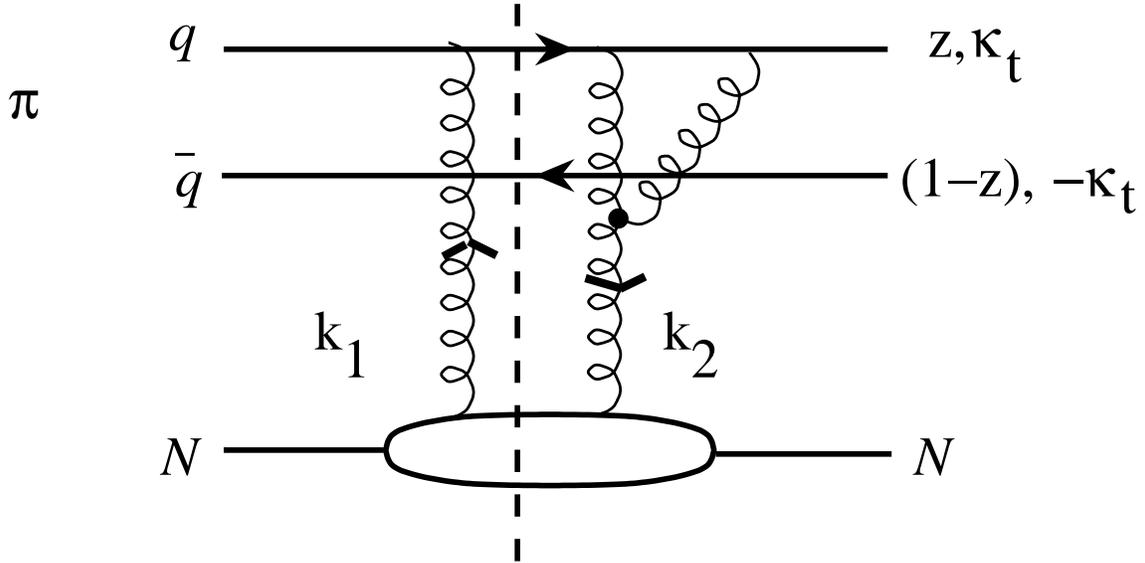, height=3.0in}
\vspace{0.3in}
\caption { Contribution to $T_{3b}$. The gluon from the two-gluon field of
the target interacts with the high momentum component of the final
$q\bar q$ pair wave function . Only a single diagram of the eight that
contribute is  shown.}
  \end{center}  \end{figure}

The term $T_{3a}$ includes the effect of the final state $q\bar q$
interaction. Fig.~6 includes the interaction of a target gluon with color flow
in the wave function of final state.

We need to evaluate only the $s-,u-$ channel cuts of the diagram (and use
Eq.~(\ref{real}) to get any necessary real part). It is useful to
define  $l_t$ as the quark transverse momentum within the pion wave
function. Then there are two kinematic regimes to consider. The first has
$l_t\ll \kappa_t, \; k_{1t}\ll\kappa_t $,
and the second $l_t^2\sim  k_{1t}^2\sim \kappa_t^2$.
We consider the former regime first, as it is expected to be
more important. In this case, we shall employ
conservation of the four-momentum to evaluate $x_2$.
Conservation of the four-momentum can be used to relate
the intermediate state (denoted by the vertical dashed line,
occurring between the emission and the absorption of
the gluons by the target in the diagram of Fig.~5\cite{torder})
of momentum $\tilde p$ with $\tilde{p}^2\equiv \tilde{m}^2$
with the intermediate state. The mass of the $q\bar q$
intermediate state is given by
\bea
\tilde{m}^2 \approx  x_1\nu -x_1\beta\nu -k_{1t}^2
\label{tilde1}
,\eea
where $\beta $ is the light-cone fraction of the
pion momentum carried by an  exchanged  gluon:
$\beta={-k_1^-\over p_{\pi}^-}={-k_{2}^-\over p_{\pi}^-}.$
Thus we arrive at the equation:
 \bea
x_1={\tilde{m}^2 + k_{1t}^2\over (1-\beta )\nu}.\label{diff2}
\eea
It follows from the requirement of positivity  of
energies of all produced particles in the
intermediate states that $0\leq \beta \leq 1$.
We can now calculate $\tilde{m}^2$ directly
in terms of
the light cone momenta of the 
$q\bar q$ pair  in the  intermediate state:
\bea
\tilde{m}^2= \left({l_t^2\over z}+{(k_{1t}-l_t)^2\over 1-z-\beta }\right)
(1-\beta) -k_{1t}^2.
\label{tilde2}
\eea
Combining Eqs.~(\ref{tilde1}),(\ref{tilde2}) we obtain
\bea
{l_t^2\over z} + {(k_{1t}-l_t)^2\over 1-\beta -z} = x_1\nu ,
\label{comb}
\eea
which, when using Eq.~(\ref{diff1})  leads to
\bea
x_1={1\over \nu}
 \left({l_t^2\over z}
+ {(k_{1t}-l_t)^2\over 1-\beta -z }\right)= {m^2_{2jet}\over \nu} + x_2,
\label{comb18}
\eea

Therefore

\bea
x_2={1\over \nu} \left({l_t^2\over z} +
{(k_{1t}-l_t)^2\over 1-\beta -z } -
{\kappa_t^2\over z(1-z)}\right).
\label{comb1}
\eea
In order for the term $T_{3a} $ to compete with $T_{1a}$ we need to have
$l_t\ll \kappa_t$, $k_{1t} \ll \kappa_t$ -otherwise $T_{3a}$ will be
additionally suppressed
by the power of $\kappa_t^2$,$\alpha_s$. These kinematics cause
Eq.~(\ref{comb1}) to yield the result:  $-x_2\propto \kappa_t^2/\nu $.

This argument can be carried out for all combinations of diagrams
represented by Fig.~5. For example, another attachment  of gluons, in
which the gluon $k_1$ is absorbed by the quark, corresponds to interchanging
$z$ with $ 1-z$, and therefore leads to the same result for $x_2$.
Evidently this result for $x_2$ is valid in the leading
$\alpha_{s}\ln \kappa_{t}^2/\Lambda_{QCD}^2$ approximation also.
Thus we consider the second situation: $l_t^2 \sim k_{1t}^2\sim \kappa_t^2$.
In this case, the  initial pion wave function contains a hard quark, and
we discuss  hard radiative correction in the next order of $\alpha_{s}$.
This is the typical situation in which there are extra hard lines, as
compared with the dominant terms, and one obtains a suppression
factor $\sim 1/\kappa_t^2$ which could be  compensated by $d^2 k_{t}$
integral. However this integral does not produce
$ \ln \kappa_{t}^2/\Lambda_{QCD}^2$ because the region of integration
is too narrow. So this contribution is at most NLO correction over $\alpha_s$.
But we restrict ourselves by the LO contribution only.

The presence of  color flow in the wave function of the
final state leads to the
interaction of a target gluon with a gluon in the wave function of the
final
state, see Fig.6. This term is suppressed by an  additional power of
$1/\kappa_t^2$. The proof of this statement repeats the same reasoning
as that explaining the suppression of the term $T_{1b}$ . It heavily uses
the WW and the Gribov representations,
discussed in  subsection A, and the  identities which follow from the
antisymmetry of the vertex for the three gluon interaction,
 the color neutrality of the pion wave function,  and the dijet final state.
In the derivation it is helpful to use the observation
that effectively $|x_2|\propto \kappa_t^2/\nu$. Evidently similar
reasoning is applicable  in computing amplitudes to leading order in
$\alpha_{s}$ and all orders in  $\alpha_{s}\ln \kappa_{t}^2/l_t^{2}$.

Repeating the same reasoning as in the estimate of the terms of
$T_{1a},T_{1b}$, and remembering that $-x_2\approx \kappa_t^2/\nu$
we achieve the estimate:
$T_3 \propto \alpha_s^2 x_1G_A(x_1,x_2,\kappa_{t}^{2})/\kappa_{t}^4$.
It is instructive to investigate whether the Feynman mechanism,  where
the leading quark (anti-quark) carries a fraction of the
pion momentum $z'$ close
to 1 but high momentum jets are formed by the action of a
final state interaction,
may compete with the PQCD description. In this case transverse momenta of
constituents $l_t$  in the pion wave function are expected to be equal
to the mean transverse momenta of partons in the non-perturbative regime.
For certainty let us model the Feynman mechanism by assuming that recoil
system is quark(anti-quark) with momentum $1-z'$ close to 0 . Within
this model we will obtain Feynman diagrams for the term $T_2$, but
with the region of integration defined by the Feynman mechanism.
A simple dimensional
evaluation of term $T_3$  due to the Feynman mechanism
within the Gribov representation
shows that it is suppressed by the powers of $\kappa_t$.
The contribution of the region
${l_t^2\over (1-z')} \ll M^2_{2jet}$ has been considered above-it
is additionally suppressed for Feynman mechanism by the restriction
of the region of integration over $z'$. Thus our  next discussion is restricted
by the consideration of the contribution of the region:
${l_t^2\over (1-z')} \geq M^2_{2jet}$.

\bea
T_3\propto {1\over \kappa_t^2} \int \psi_{\pi}(z', l_t^2)
{1\over M^2_{int}-M^2_{2jet}}{(l_t)^2\over (1-z')}
d^2 l_t dz'
\label{Feynmanmech}
\eea
In the above formulae we use Brodsky-Lepage convention for the
definition of wave functions and retain terms maximally singular when
$z'\to 1$. Power counting is simple: the factor
${l_t^2\over (1-z') \kappa_t^2}$ is from the gluon exchange in the
final state. The factor $l_t^2/(1-z')$ is singular when $z'\to 1$. It originates 
 from the quark vertexes accompanying the propagator of the gluon
exchanged in the wave function of final state. Here $1/(1-z')$ follows
from a transition when a fraction of the pion momentum carried by quark
tends to 0. The factor
${M^2(2jet)-l_t^2/z'(1-z')}$ in the denominator is due the fermion propagator
adjacent to the hard gluon exchange in the wave function of the final
state.Here $M^2_{int}\approx {m^2_{rec}+l_t^2\over 1-z'}$
is the mass of an intermediate state, and  $m_{rec}^2$ is the invariant  mass
of the recoil system in the Feynman mechanism.
In the region of integration $1-z'\ll l_t^2/M^2_{2jet}$
one may neglect by $M^2(2jet)$ in the denominator as compared to
$l_t^2/(1-z')$. So one obtains : $T_2\propto {1\over \kappa_t^2}
\int \psi_{\pi}(z', l_t^2) d^2 l_t dz'$. In this case, another
factor of $1/\kappa_t^4$ arises from the
integration over $z'$. Hence we have found that the Feynman mechanism
is a higher twist correction to the PQCD contribution. The Feynman mechanism is
further suppressed by the requirement of a lack of collinear to pion
momentum radiation-see the discussion below.

\subsection{Gluon admixture to the wave functions of
initial and final states - $T_4$}

The Feynman diagram corresponding to Fig.~7
contains the time ordering corresponding to  the $q\bar q g$ configuration
in the pion wave function interacting with the quarks in the final state.
In taking the imaginary  part of the amplitude, the intermediate
state must contain a hard on-shell quark and a hard on-shell gluon. But
such a state cannot be produced by a soft almost on-shell quark in the
initial state, so there is an additional suppression factor, caused by
the rapid decrease of the non-perturbative pion wave function with
increasing quark virtuality. This factor is greater than a power of
$\kappa_t^2$. One may also consider the case when the transverse momenta
of quarks in the pion wave function are large enough to use PQCD. Then the
large virtuality of the quark introduces a suppression
factor of  ${1\over \kappa_t^2 l_t^2}$, with at least one power of
${1\over \kappa_t^2}$ arising from the quark line for the transition
$q\to qg $ and another factor of
${1\over {\l}_t^2}$ arising from the pion wave function. There are additional
factors:
${1\over \kappa_t^2}$ arises from the   hard fermion line, and
${1\over \kappa_t^4}$  from the application
of Ward identities and the condition: $x_1 \nu,x_2 \nu \propto \kappa_t^2$.
A factor of $\kappa_t^2 l_t^2 $ is present in the numerator, with
$\kappa_t^2  $ originating  from the
vertices in the WW representation and $l_t^2$
from the integration over quark momenta in the pion wave function.
All in all this amplitude is suppressed by the factor $l_t^2/(\kappa_t^2)^3$.
Another case occurs when $l_t^2\propto \kappa_t^2$. Then  this diagram
will be suppressed as compared to
$T_1$ at least by one power of $\alpha_s$ without the large factor
$\ln\kappa_t^2/\Lambda_{QCD}^2$.  But here we restrict ourselves to
the analysis of LO corrections.

\begin{figure}[t]
\begin{center}
\epsfig{file=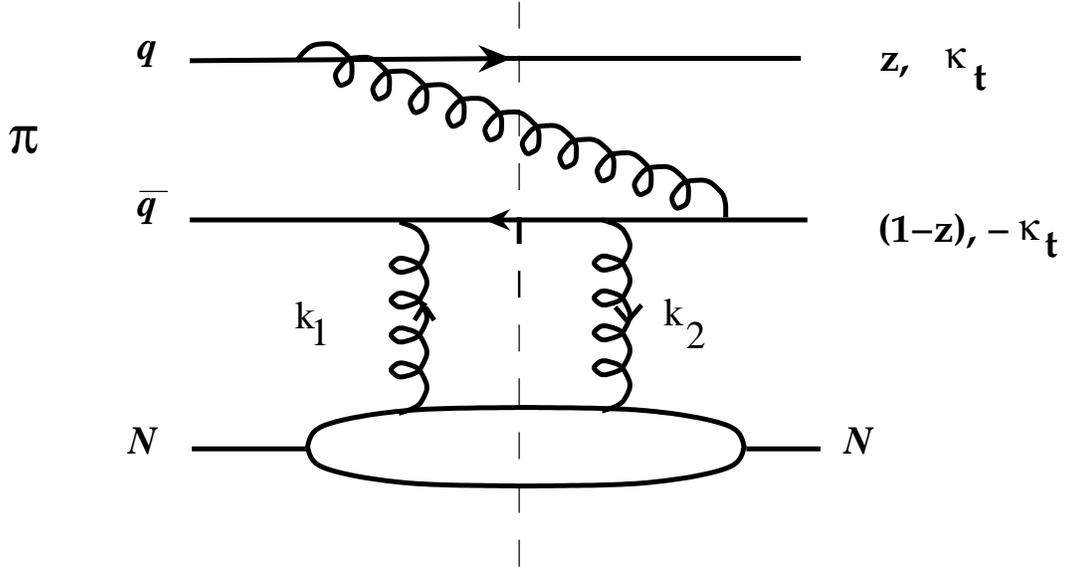, height=3.0in}
\vspace{0.3in}
\caption{ A time ordering that contributes to $T_4$.
The $q\bar{q} g$ state interacts with the target.
Only a single diagram of the eight where a gluon interacts with quarks
in a pion fragmentation region that contribute is  shown.}
 \end{center}
  \end{figure}
Similar reasoning helps to prove that the contribution of diagrams in Fig.~6
is suppressed by a factor $l_t^2/\kappa_t^2$ as compared to that in  Fig.~1.
This is the power-type suppression if the 
pion wave function is non-perturbative,
and  may be a NLO $\alpha_s$ correction if the
perturbative high momentum tail is included in the pion wave function.

\begin{figure}\begin{center}
\epsfig{file=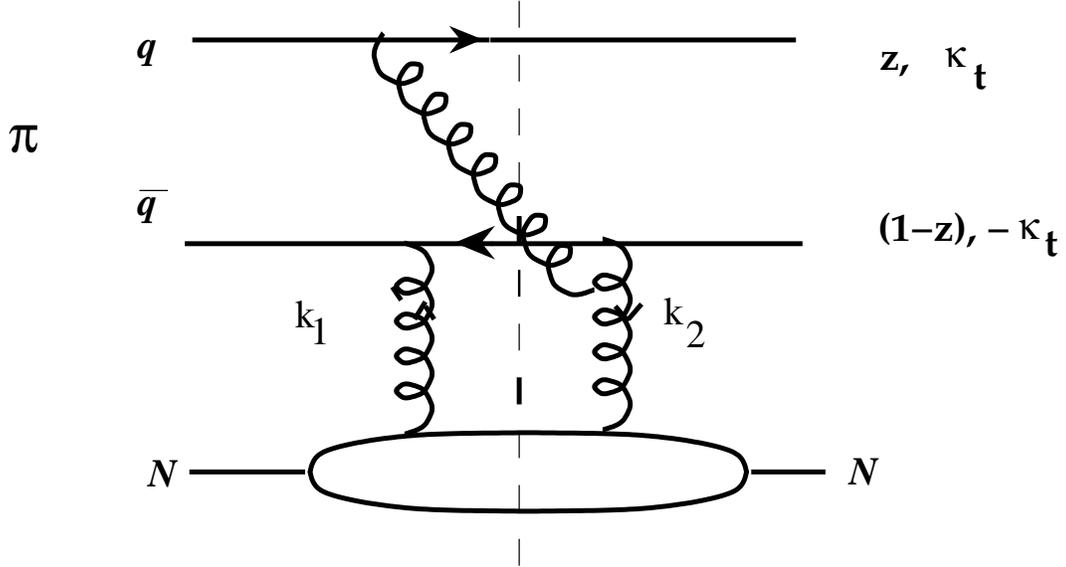, height=3.0in}
\vspace{0.3in}
\caption{A contribution to $T_{4b}$.
The target gluon absorbs a gluon of pion wave function.
Only one diagram of the eight that occur is shown.}
 \end{center}   \end{figure}

Another contribution to $T_4$ arises from the sum of Feynman diagrams in
which the gluon exchange between the $q$ and $\bar q$ in the beam occurs
during the interaction with the target, see Figs.~8,~9, and 10.
The naive expectation is that such terms, which amount
to having a gluon exchanged during the very short interaction time
characteristic of the two gluon exchange
process occurring at high energies, must be very small indeed.

The intent of this sub-section is to use the analytic properties of the
scattering amplitude to show that $T_4$ is negligible. Instead of
calculating the sum of the imaginary parts of all of the amplitudes,
we will prove that this sum vanishes  by analyzing analytic properties of
the important diagrams.  Each considered
diagram contains a product of an intermediate-state quark and anti-quark
propagator. At high energies, these propagators are controlled by the terms
of highest   power of $x_1\;2p\cdot p_{\pi}=x_1 \nu$, and (as to be shown)
have poles in the complex $x_1$ plane  which are located on one side of
the contour of integration. The sign of the term containing $(\nu)$
in each propagator unambiguously follows from the directions of pion
and target momenta. If we can show that the typical integral is of the form

\bea
\int dx_1 \frac{1}{(\alpha\;x_1\nu-a+i\epsilon)(\beta\;x_1\nu-b+i\epsilon)},
\qquad\alpha,\beta>0\label{disp}\eea
the proof would be complete.

\begin{figure}\begin{center}
\epsfig{file=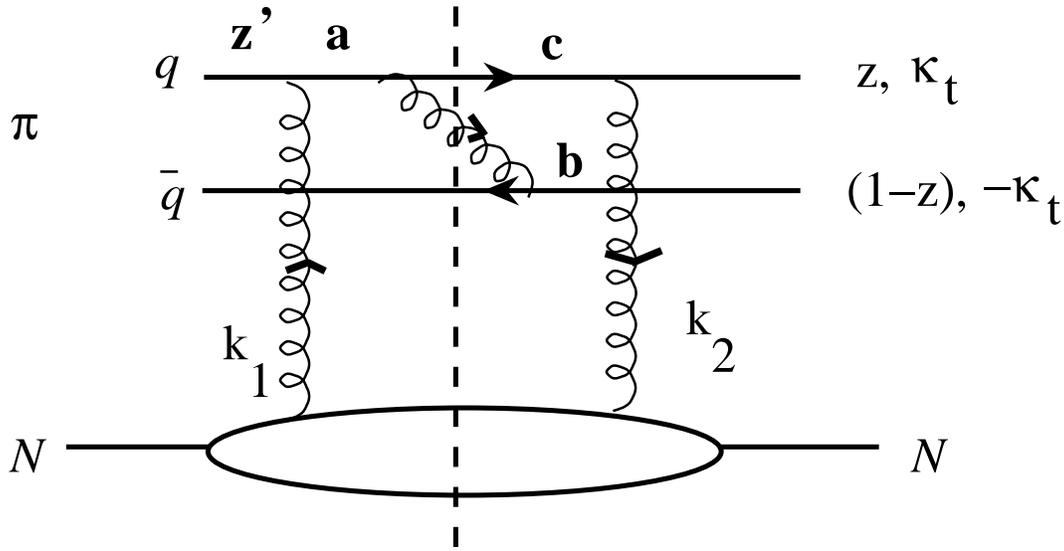, height=3.0in}
\vspace{0.3in}
\caption{A contribution to $T_{4c}$.
The quark absorbs a gluon of momentum $k_1$, exchanges a gluon with
the anti-quark, and then emits a gluon of momentum $k_2$. Only one
diagram of the eight that occur is shown.}
 \end{center}   \end{figure}

We now consider the Feynman graphs, starting with Fig.~9.  Once again we
compute the imaginary part of the graph and consider the intermediate state
as being on the energy shell. The propagator for the line (a) has the factor

\bea (k_1+z'p_\pi)^2-m_{q}^2 =z'x_1\nu +\cdots
,\label{pa}\eea
while that of the near-mass-shell line (b)  is independent of $x_1$,
because the quark momenta
in the final state and in the pion wave
function are not connected with the target momentum. The
propagator of line (c) has the factor
\bea
(k_2+q_1)^2-m_{q}^2=x_2\;z\;\nu+\cdots=x_1\;z\;\nu+\cdots.
\label{pc}\eea
Here $q_1$ is the momentum of the jet $(z,\kappa_t$) and $\cdots$
denotes the terms which are independent of $x_1$.
The last equation is obtained from
using Eqs.~(\ref{diff1},\ref{x2}).
The results (\ref{pa}-\ref{pc}) show that
the diagram of Fig.~9 takes on
the mathematical form of the integral
(\ref{disp}). Thus this term vanishes.
\begin{figure}\begin{center}
\epsfig{file=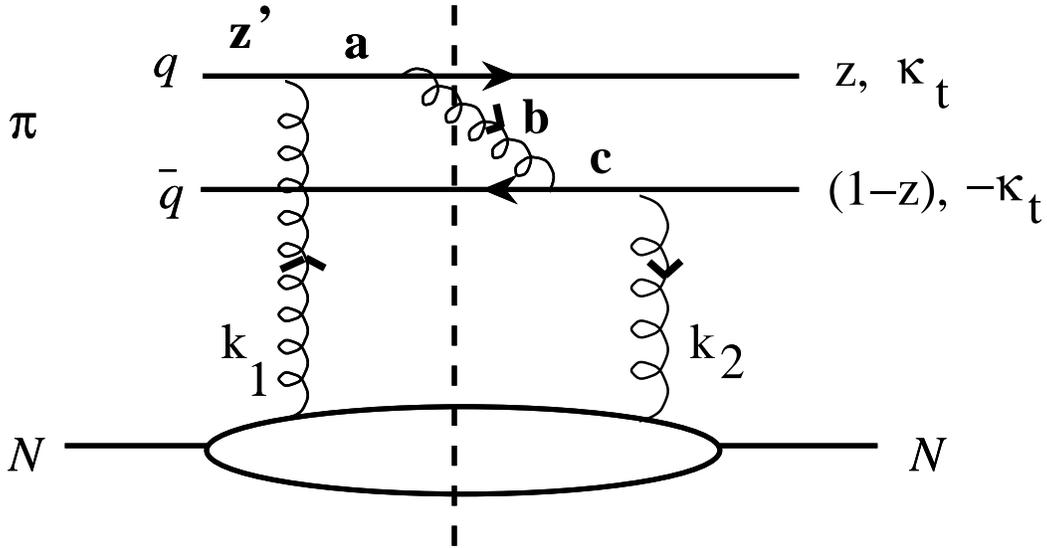, height=3.0in}
\vspace{0.3in}
\caption{ Another contribution to $T_4 - T_{4d}$.
The quark absorbs a gluon of momentum $k_1$, exchanges a gluon
   with the anti-quark, and the anti-quark
   emits a gluon of momentum $k_2$. Only one
   diagram  of the four that
   contribute is shown.}
 \end{center}   \end{figure}

We also consider the diagram of Fig.~10. In this case there are three
propagators (a), (b), (c) that  have a term proportional to $x_1 \nu$, but
the coefficients are not all positive.
The  propagator factor for line (a) is given by
\bea
(x_{1}\;p_\pi+k_1)^2=x_1z'\nu+\cdots,\eea
while that of line (c) is given by
\bea
(k_2+q_2)^2-m_{q}^2=x_2(1-z)\nu+\cdots=(1-z)x_1\nu +\cdots
\eea
At the same time, the coefficient multiplying $x_{1}$
in the propagator (b) (gluon production) has no definite sign. Thus
for this diagram the integral over $x_1$ does not vanish.
The presence of an additional gluon in the intermediate state means
that $x_{1} \propto x_{2}$. This and the  use of the
Weizsacker-Williams representation allow us to conclude that this
diagram is suppressed by a power of $\kappa_{t}^2$ as compared to $T_{1}$.
Similar logic can be applied to any of the diagrams contributing to $T_4$.
The physical idea that the intermediate $q\bar q$
state does not live long enough to exchange a gluon
is realized in  the ability to close the contour
of integration in the upper half plane
or in the suppression by the power of $\kappa_{t}^2$.
The analyticity of the scattering amplitude in the upper half
is a consequence of causality. Thus the physical and mathematical
ideas behind the vanishing of $T_4$ are basically equivalent for all
diagrams at high enough beam energies.

We conclude this section with a brief summary.
We analyzed the leading diagrams for the pion
dissociation into two jets and found that the hard dynamics-amplitude
$T_1$ determines this process, with the initial state wave function
determined by the  hard gluon exchange  diagram. In the next section we
shall show that amplitudes $T_2,T_3$ are strongly
suppressed by the requirement of the lack of radiation in the final
state. This is because in the lowest order in
$\alpha_s$
these amplitudes correspond to the propagation  of a
nonperturbative q$\bar q$
dipole or a q $\bar q$ g tripole over  large longitudinal distances.

While we were revising the manuscript in response to referee's questions,
a  paper \cite{Braun} appeared which claimed that, if one includes terms
beyond the leading-logarithm approximation in
$\alpha_s \ln x$,  factorization 
does not hold and  end-point singularities break
collinear factorization. Our calculation shows that such problems are not
present in the leading order approximation which keeps terms  leading  order
in
$\alpha_s \ln {k_t^2\over \Lambda_{QCD}^2}$.
The use of our light cone  gauge, where the
asymptotics of
the fermion propagator has no infrared singularities in the hard regime,
makes the separation of scales-QCD factorization- rather straightforward.
Moreover in this gauge, terms $\sim\ln{1/x}$ in the hard regime are related
to the exchange by the gluon in the multi-Regge kinematics only
\cite{Gribov}. As a consequence of the QCD factorization theorem terms
$\sim$ $\ln 1/x$ are included in
structure functions of the target. On the contrary, if one uses the
standard gauge $A^+=0$ or Feynman gauge, the cancellation of infrared
singularities occurs if one includes the renormalization factor
arising from the hard Fermion propagator,  cf. discussion in
Ref.~\cite{BL}. Taking these terms together, the resulting amplitude
contains no end-point singularity, and the factorization holds.

\section{Post-selection of the projectile wave function by fixing the
final state}

A specific feature of the processes considered here is that the
restriction on the composition of the final state selects a rather
specific initial configuration of the projectile hadron. Following
S.Nussinov (to be published) we denote  such measurements as
post-selection. In our case, the initial and final state wave
functions are built at large longitudinal distances $\propto 1/2m_Nx$,
so there is plenty of time for  radiation to occur.  But our
definition of the final state forbids radiation collinear to the direction
of pion momentum. In particular, the relation between the
transverse momentum
of jet $\kappa_t$ and the mass of  the diffractively produced system,
$M^2(2jet)=\kappa_t^2/z(1-z)$, is natural only  for a $q\bar q$ final
state. Thus any contribution due to processes for which such radiation
is kinematically and dynamically permitted must be suppressed by the
powers  of Sudakov-type form factors and by a $w_2$ form factor.  The
first step is to analyze how the trigger for the two jet state
restricts  the composition of the final state.

\subsection{Three jet production.}
\label{3jets}
A question arises\footnote { We thank J.Bjorken, and D.Soper for asking this
question.} whether the trigger used in Ref.~\cite{danny} allows the
separation of production by  a  $q\bar q $ state from  the  production
by a 
$q\bar q g$ state (and more complicated states containing relatively
soft  partons) as the source of the observed dijets, and whether the
presence of such states may change the $\kappa_{t}$ dependence of
cross  section. Such states are certainly important in  inclusive
diffraction in DIS at HERA. A typical diagram  corresponding to such
a process is presented in Figure 11.

\begin{figure}\begin{center}

\epsfig{file=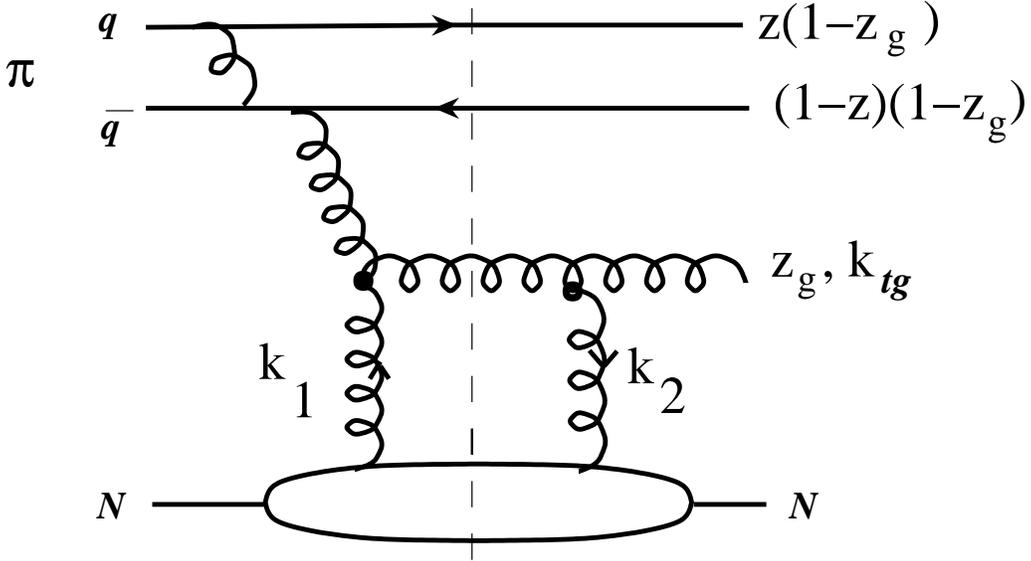, height=3.0in}
\vspace{0.3in}
\caption{A typical diagram corresponding to production
of large mass $q\bar q g$ state.}
\end{center}   \end{figure}
An analysis of kinematics shows that the
virtuality of the gluon interacting with the target is
$\approx -\frac {(1-z_{g})M_{2jet}^2+k_{tg}^2}{z_{g}}$
where $z_{g}$ is the fraction of the 
pion momentum carried by a gluon in the final
state and $k_{tg}$ is its transverse momentum, and as usual, the mass
of the final dijet system is given by
$m_{f}^2=M_{2jet}^2=\kappa_t^2/z(1-z)$.
The mass of a $q\bar q g$ system in the final state is given by
$M_{3jet}^2=\frac{M_{2jet}^2+k_{tg}^2}{(1-z_{g})} +k_{tg}^2/z_{g}$.
If $z_{g} \leq k_{tg}^2/M_{2jet}^2$  then
$M_{3jet}^2\geq 2  M_{2jet}^2 $, and therefore the $q\bar q g$ state is
distinguishable from the two-jet state  by the relation between $\kappa_t^2$
and the total mass. In this kinematics, the cross section depends on
$\kappa_{t}$ as $ \alpha_{s}(k_{t}^2)^3
\frac{(\alpha_{s}(k_{tg}^2) x_{1}G(x_1,x_2,k_t^2))^2} {k_{t}^8}$.
Beyond this kinematics, the  cross section is additionally suppressed by the
factor  $\propto ({z_{g}M_{2jet}^2})^{-2}$ .
Thus a contribution of such configurations is either dynamically suppressed,
or distinguishable from the contribution of a $q\bar q$ configuration.
Note also that the feature that $k_{tg}$ is  small would not lead to
the cancellations needed for color transparency phenomenon to occur, and
therefore the A-dependence would differ from what is observed.

It is important to be able to distinguish dijet production from the
soft diffraction into a 
hadronic state having a total  mass $M^2_{diff}$ and  containing  leading
twist dijet production with mass $M_{2jet}^2$. Evidently
$M_{2jet}^2\leq M^2_{diff}$.
Cross section of such diffractive processes are often described as
the hard diffractive mechanism of Ingelman-Schlein\cite{IS}. In
this mechanism one considers a hard scattering of a parton belonging
to the projectile $\pi$ with a light-cone fraction $x_{\pi}$ off a
parton belonging to a diffractive parton density (effective Pomeron)
with a light-cone fraction $\beta x_{I\!\!P}$.

Here $1-x_{I\!\!P}$ is the nucleon momentum in the final state,
cf \cite{IS}. It follows from the above definitions and energy-momentum
conservation law that
$R^{-1}\equiv {M_{2jet}^2  \over M_{diff}^2}=x_{\pi}\beta$
Thus $R^{-1}$ is the fraction of the total mass of 
the diffractive state
carried by two jets.
We are interested in the paper in the limit  when
the mass of a diffractively produced hadronic state is carried mostly by 2 jets
i.e. when $ R\equiv R_0\sim 1$.
In this case the cross section for the production of dijets with $R\geq R_0$
(where $R_0$ being close to 1 is determined by the accuracy of
measuring a fraction of the pion momentum carried by  two jets) is
\begin{equation}
{d\sigma\over d\kappa^2}\propto {1\over M_{diff}^2}\kappa^{-4}
\int_{R_0}^1dx_{\pi}\int_{R_0/x_{\pi}}^1 d\beta   (1-x_{\pi})^2
f_{I\!\!P}(\beta).
\end{equation}
Here a factor $\kappa^{-4}$ is the usual $\kappa$ dependence of the
cross section of the hard two parton collision, ${1\over M_{diff}^2}$
is the usual Pomeron flux factor,  the factor $(1-x_{\pi})^2$ is
the parametrization of
the parton density in
the pion at $x_{\pi}\to 1$, $f_{I\!\!P}(\beta)$ is the diffractive
structure function. Taking $f_{I\!\!P}(\beta) \propto (1-\beta)$
for $\beta \to 1$ (see e.g. \cite{H1}) we obtain
$d\sigma/d\kappa^2 (R\geq R_0)\propto (1-R_0)^5/\kappa^6$.
The factor $(1-R_0)^3$ arises from the integration over $x_{\pi}$
and factor $(1-R_0)^2$ is from the integration over $\beta$.
The additional factor ${1\over \kappa_t^2}$ is because in the
discussed region $M_{diff}^2\approx M_{2jet}^2$.
There is also a contribution of the Coherent Pomeron \cite{CFS93}
corresponding to $\beta=1$. It leads to a similar suppression
as a function of $R_0$: $\propto (1-R_0)^4$.

Hence we conclude that the  importance of the discussed leading twist
mechanism  as compared to the the exclusive dijet production term
which is $\propto  \kappa^{-8}$ depends on the degree of exclusiveness
of the experiment. The experiment \cite{danny} imposes the condition that
$M_{2jet}^2 = M_{diff}^2$. Such a selection should have been rather
efficient since due to the acceptance of the detector a condition
was imposed that all produced pions should have momenta larger
than a minimal one.

\subsection {Suppression of the final state interaction}

Let us consider the important consequences of the formulated above
restrictions on the composition of the final state. The existence of
the  term $T_3$ displayed in Fig.~5, caused Jennings \& Miller
\cite{jm} to worry that the value of ${\cal M}_N$ might be severely
reduced due to a nearly complete cancellation.  However, we shall explain
here that this term
as well as the  term $T_2$ are
strongly suppressed in QCD as compared to the
naive PQCD calculation explained in Sect.II.
This suppression is the only way to resolve an evident contradiction:
the kinematic restriction on the final state discussed above  forbids
radiation collinear to the pion momentum, but such radiation naturally
arises in a hard collision, or from the presence of a significant
gluon admixture in the non-perturbative pion wave function
when nonperturbative q$\bar q$ and more complicated configurations
propagate large and increasing with energy longitudinal distances.
It is convenient to represent this suppression as a product of two
factors: $w=w_1 w_2$. The first factor accounts
for the well understood suppression  of  the collinear initial state
radiation in the scattering of a target gluon off a low $k_t$ quark.
Remember that by definition a non-perturbative pion wave function does
not include a PQCD high momentum tail. Accounting  for the LO QCD evolution
will not change this conclusion in the LO approximation over parameter
$\alpha_s\ln {k_t^2\over \Lambda^2_{QCD}}$. In the case of the amplitude
$T_3$ the diagrams where the hard gluon exchange is present both in the
initial and the final states are potentially dangerous.
However,  this contribution into $T_3$
is suppressed by the power of $\alpha_s$ as compared to the amplitude $T_1$.
Similar radiation is permitted for the processes described by
the amplitude $T_3$ because tripole q$\bar q$g propagates large
longitudinal distances. This  radiation in the direction of pion momentum
carries a finite fraction of pion momentum which is significantly
larger than that for wee hadrons which are products of the jet
fragmentation. These wee hadrons carry
$\propto {few~ m_{\pi}\over M(2jet)}$ fraction of pion momentum.
The $w_1$ form factor suppression for the amplitude $T_3$
is given by the square of Sudakov type form factor:
$S^2(\kappa_t^2/l_t^{2})$; one form factor arises
for each collision with a target gluon. Here $l_{t}$ is the transverse
momentum of a quark within the non-perturbative pion wave function.
This form factor is a square root of the form factor of quark\cite{sudff}
because the radiation of gluons off the final quark is included in the
definition of the final state. In the  light cone gauge:
\bea
w_1=S^2(\kappa_{t}^2/l_{t}^{2})=
 \exp(-4/3\frac{\alpha_s}{4 \pi } \ln^2 \kappa_{t}^2/l_t^{2}),
\eea
where, using   the leading log approximation, we replaced  $k_{1t}$
by $\kappa_t$ in the argument of the Sudakov form factor.
Recall  that the form factor $S^2$ is exactly the Sudakov form factor
which enters in the Dokshitzer-Gribov-Lipatov-Altareli-Parisi
evolution equations as a coefficient of the $\delta(x-1)$ term, see the
discussion in Ref.\cite{DDT}.

$w_1$ form factor should be practically the same for both amplitudes
$T_2$ and $T_3$.  This is because  of total transverse momentum  of
the system of high $k_t$  quark(antiquark) + high $k_t$ gluon in the
amplitude $T_3$ is small and controlled by the nonperturbative pion
wave function. Therefore in the collision with a target gluon this
quark(antiquark) - gluon system radiates in the direction of pion
momentum as  a single quark(antiquark).

The second  suppression factor, $w_2$, accounts for the dependence of
the form of the effective QCD Lagrangian and  appropriate degrees of
freedom on the resolution. As a result of partial conservation
of axial current  $f_{\pi}$ is independent on the resolution-invariant on
renormalization group transformations. At the same time important
degrees of freedom depend on the resolution: dominant degrees of freedom
in hard processes are bare quarks. On the contrary, in the non-perturbative
regime because of the effects of spontaneously broken chiral symmetry
dominant degrees of freedom  are constituent quarks,
pseudogoldstone modes -pions, various condensates of quark, gluon
fields, etc. Remember that
for the average-sized configuration  of a hadron, approximately half  of
the pion momenta is carried by gluons. (Within a constituent quark
model renormalizability of QCD is usually accounted for by introducing
effective mass and effective interaction for the constituent quarks
and $f_{\pi}$ is calculated in these models in terms of the constituent
but not bare quarks. The
evaluation of $f_{\pi}$ within a bag model should include a prescription how
to treat the nonperturbative volume energy density - bag surface.
Within the Weizecker-Williams approximation this energy is equivalent
to the gluon cloud   in the light cone wave function of a pion .)

In the case of interaction of local current $f_{\pi}$ is calculable
in terms of the distribution of bare quarks which accounts for both
nonperturbative and perturbative effects \cite{BL}. In this case
the suppression factor $w_2$ tends to one for $k^2_0\to \infty$ since in
this case only short distance perturbative degrees of freedom survive.
So amplitude $T_1$ containes no additional suppression factor $w_2$. On the
contrary in the amplitudes $T_2,T_3$ the structure of constituents in the
non-perturbative pion distribution is resolved as a result of a large
time interval between two consequent hard collisions of pion
constituents off target gluons. A hard collision of the pion's constituents
with target gluons (with momentum transfer ($k_0^2\ll \kappa^2$))
frees some gluons from the pion wave function,  and emitted radiation is
collinear to the pion momentum. This radiation is, however, forbidden by
the restriction on the final states discussed above. Thus the
selection of a component of the pion wave function of a size
determined by non-perturbative QCD, leads to an additional factor $w_2$ which
suppresses this contribution to the scattering  amplitude.

We may estimate the factor $w_2$ using models as follows:
\bea
w_2=\frac{\int_{0}^{1} dz_1\int_0^1
  dz_2 (1-z_1-z_2)^n \theta (\delta -1+z_1+z_2)
\theta(1-z_1-z_2)} {\int_{0,1} dz_1 dz_2 (1-z_1-z_2)^n \theta(1-z_1-z_2)}
=(n+2)(\delta)^{n+1}.
\eea
Here $z_{1,2}$ are the fractions of the pion momentum  carried by jets
in the final state, and $\delta$ is the experimental uncertainty in
the fraction of pion momentum carried by high $\kappa_t$ dijets.
Within the democratic chain approximation, which reasonably describes the
$Q^2$ and $x$ dependencies of hard exclusive processes, the value of $n=1$
for the component which contains $q\bar q$ and a valence gluon.
Another estimate of $n$ can be obtained in  the constituent quark model
assuming, for simplicity, that each constituent quark consists of
a bare quark and one gluon. In this case we are effectively dealing with
a four particle system and hence $n\geq 2$. The factor $w_2$ gives a
significant numerical suppression for all the  models.
No such suppression appears in the amplitude
$T_1$ because the radiation of a gluon with small transverse momentum from a
highly localized (e.g. size  $1/\kappa_t$) color-neutral
quark configuration  is suppressed by a  power
of $\kappa_t^2$. (Note that in the case of the final state interaction,
the non-perturbative pion wave enters at 0 inter-quark transverse distances.
This means that  gluon radiation  with transverse momenta,
$k_{tg}^2\leq l_t^2$, should be suppressed. For  gluon radiation with
larger transverse momenta there is no restriction.)
Amplitudes $T_2,T_3$ correspond to the propagation at large
longitudinal distances of ``large'' size  q$\bar q$ pair whose
transverse size is controlled by non-perturbative pion wave function
The net result of all of this is that amplitudes $T_2,T_3$  and
Figs. 3,5,6 can be neglected  and we shall therefore ignore the
amplitudes $T_2,T_3$.

We also observe that the contribution of Feynman diagrams in which
the skewed distribution is modeled by the scattering of an on-shell quark
or gluon is suppressed even further by the square of the Sudakov form
factor, in addition to form factors $w_1,w_2$.

\subsection{Non-leading order approximation to the dipole-target interaction}

 In the previous section we used the leading
$\alpha_s \ln\kappa_t^2/\Lambda_{QCD}^2$ approximation (for the case
$k_{1,2t}^2\ll \kappa_t^2$) to estimate diagrams. To go beyond this,
one should take into account the contribution of the region
$k_{i,t}^2\approx \kappa_t^2$ to obtain
the $q\bar q$  dipole-target interaction.
The related contribution to dijet production is
 denoted as $T_5$. To some extent this
contribution has been discussed above,  in relation to the amplitude
$T_4$. For the amplitude $T_5$, quarks in the wave function of the initial
pion  have small transverse momenta $l_t$, and there is no hard
interaction between the $q$ and $\bar q$ in the initial or final states,
see Fig.~12.
\begin{figure}\begin{center}

\epsfig{file=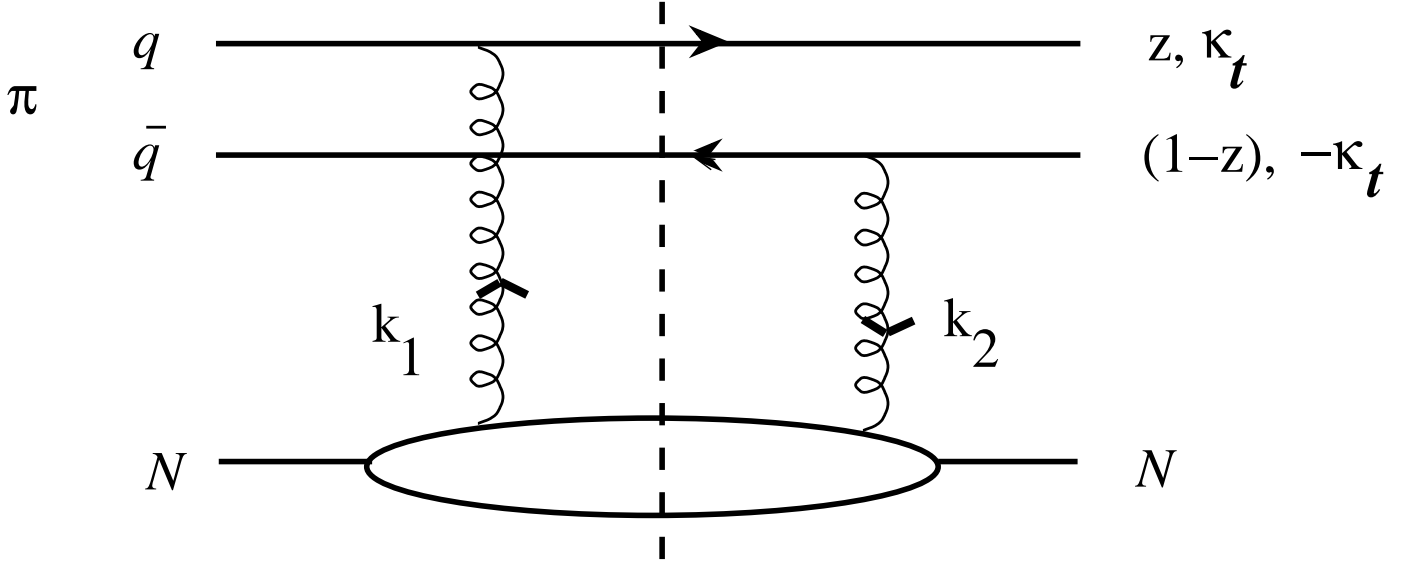, height=3.0in}
\vspace{0.3in}
\caption{ $T_5$, Hard color flaw diagram.}
\end{center}   \end{figure}

Let us outline various phenomena relevant for the smallness of
this contribution as compared to $T_1$.

\begin {enumerate}

\item The amplitudes given by  diagram  Fig.~12 correspond to  two hard
collisions occurring
at the different space-time points where color abruptly changes
the direction of its motion. The longitudinal distances (time interval)
between both hard collisions  are large,
$\approx 2E_{\pi}/\kappa_t^2$,
so that the emission of gluon radiation is permitted.
The difference in impact parameters
characterizing both hard collisions is
$\approx {\pi \over 2 {\l}_t}$,
where $l_t$ is the transverse momentum of quark in the
non-perturbative pion wave function. Thus collinear gluon radiation
in both collisions is not canceled,
and  forward gluon bremsstrahlung (precluded
in $T_{1}$ by the localization of color in transverse space) occurs.
Such radiation effects are precluded by our choice of final states for which
 radiation collinear to pion momentum is absent. Thus the ratio
$T_5/T_1$ is suppressed (as explained in the previous subsection) by the
square of a Sudakov-type form factor: $S^2(\kappa_{t}^2/l_{t}^{2})=
\exp(-4/3\frac{\alpha_s}{4 \pi} \ln^2 k_{1t}^2/l_t^{2})$.
Including the effects of the
Sudakov type form factor  leads to an increase of the effective value
$l_t$ with an increase of $k_{1t}$,
and therefore  will change kinematics of this
diagram in the direction to that for the term $T_1$.

\item Final states with additional gluons collinear to pion momentum
are initiated
by the components of the pion wave function which are close to the average.
The significant probability for such processes follows from the fact that
the gluons carry $\sim 1/2 $ of the pion momentum. Thus the selection
of the 
component of a pion wave function of average size, but having no gluons,
leads  again to the appearance of the additional suppression factor
 $w_2$.

\item As the consequence of the rapid decrease  of the
non-perturbative pion wave function with $l_t^2$  for the diagrams of
Fig.~12, the target gluon actually has  a negative value of
$x_2$. To see this, we apply Eq.~(\ref{comb1})  to the
situation when the transverse momenta of quarks in the initial state
$l_t$ are much smaller than $\kappa_t$, so that
$k_{1t}\sim \kappa_t,k_{2t}\sim -\kappa_t$.  In this case
\bea
x_2= {1\over \nu}
\left[{\kappa_t^2\over 1-\beta -z }-{\kappa_t^2\over z(1-z)}\right].
\label{rx2}
\eea
Here $\beta$ is the fraction of the pion momentum carried by 
a target gluon.
To satisfy the condition $x_2>0$ one needs
\bea
1-z > \beta > (1-z)^2, \;{\rm or}  \; z > \beta > z^2.
      \label{restr}
\eea
This  condition  cannot be satisfied within the leading $\log1/x$
approximation where $\beta \ll 1 $.  Moreover,
 Eq.~(\ref{rx2}) implies  that $-x_{2}\nu\propto \kappa_{t}^2$,
except for a narrow interval in $\beta$ near $\beta=(1-z)^2$.
The contribution of this narrow interval is suppressed
by the small length of this interval. (The end point
$z \sim 0, z\sim 1$  contributions  are suppressed by the pion wave
function.)  In the non-perturbative regime- within the parton model- the
contribution of the $x_2\leq 0$ region cannot be expressed
in terms of the unintegrated gluon density, and it is suppressed by power of
$1/\kappa_t^2$ \cite{Feynman}. On the contrary, within the leading $\log 1/x$
approximation such a strong suppression is substituted by a slower one
whose value depends on the ratio of $\kappa_t^2$ to the square of the
mass  of the  recoil
system (the wave function of a 
target where a high $\kappa_t$ gluon is removed). This
ratio is   $\kappa_t^2/\beta \nu$.
Evidently, the contribution of small $\beta$ is suppressed by
the rapid  decrease of structure function with $\kappa_t^2$. This
cutoff
 is significantly stronger than that for the term $T_1$  because of
smaller transverse momenta of target gluons.

\end{enumerate}

Thus we conclude that the  selection of the final state leading to the
post-selection of the initial state shows that approximation of
having a point-like pionic configuration
 in the final state should be valid even beyond the leading order
approximation used in this paper. Therefore formulas representing 
a hard
process as the convolution of a pion wave function with the interaction
cross section\cite{fms93} should be valid beyond the LO approximation.

It has been suggested in \cite{nnn} that gluon scattering off on-mass
shell quarks and gluons in the target wave function, see Fig.~12, is
the dominant process
for the diffractive dijet production by a  nuclear target. We explained
above that in QCD this amplitude is suppressed as compared to the term
$T_1$ by the product of the square of Sudakov type form factors
and the form factor $w_2$. So we will neglect now and forever
 the amplitudes presented in Fig.~12 .

\section{ Amplitude for $\pi N\to N JJ$--Evaluation
of the Dominant Term }

Let us consider  the forward ($t=t_{min}\approx 0$)
amplitude, ${\cal M}$, for coherent dijet production on a nucleon
$\pi N\to N JJ$ \cite{fms93}:

\bea
{\cal M}(N)=\langle f(\kappa_t,z),N' \mid
\widehat{f}\mid \pi,N\rangle,
\label {matel}\eea
where $\widehat{f}$ represents the interaction with the target nucleon.
The initial $\vert \pi\rangle$ and final $\vert f (\kappa_t, z) \rangle$
states represent the physical states which generally involve all
manner of multi-quark and gluon components.
For
large enough values of $\kappa_t$, the result of calculations can
be represented in a form in which only the $q\bar q$ components of the
initial pion are relevant in Eq.~(\ref{matel}).
This is because we are considering a coherent nuclear process which leads
to a final state consisting of a quark and
anti-quark moving at high relative transverse momentum.

We showed above  that  the  dominant (in powers of $s$ and $\kappa_t^2$)
Feynman diagrams will be expressed in terms of the light-cone wave function
of the pion.  We therefore begin  by calculating the high
transverse momentum component of the pion wave function. The
non-perturbative component  of the light-cone pion wave function is
represented by $\mid \pi\rangle$, and  the high momentum components can be
treated as arising from the following approximate equation\cite {BL}:
 \bea
 \mid \pi_{q\bar q}\rangle =
G_0 V_{eff} \mid \pi\rangle,
\label{pieq}
\eea
where $G_0(p_t,y)$ is the non-interacting
$q\bar q$ Green's function for  $p_t^2\gg m_q^2,m_{\pi}^2$
\bea
\langle p_t, y\mid G_0 \mid p'_t,y'\rangle=
{\delta^{(2)}(p_t-p_t')\delta (y-y')\over
- {p_t^2 \over y(1-y)}},
\label{g0p}\eea
in which  $m_q$ represents the quark mass,
$y$ and $y'$ represent  the
fraction of the longitudinal momentum
carried by the quark; and the relative
transverse momentum between the quark
and anti-quark is $p_t$. The  complete
effective interaction, obtainable  in principle from
PQCD, implicitly includes the effects of all
Fock-space configurations.

The evaluation of the graphs corresponding to Fig.~1 consists of two parts.
As a first step, the  Feynman diagram  of Fig.~1 can be rewritten as a product
of a high momentum component of a light cone pion wave function with the
amplitude for the scattering of a quark-anti-quark dipole by a target. So,
we need to know the relevant part of the pion wave function. Secondly,
we need to determine the interaction with the target (here with the
gluon field of the target) which causes the pion to  dissociate
into a $q\bar {q}$.

\subsection {High momentum component of  the light cone pion wave 
  function\protect\footnote{An
 early version of this subsection which
    did not include discussion of renormalization effects has appeared
    in \protect\cite{hallertext}.
}  
}

%Much of this section has appeared previously\cite{hallertext}. We
%include this material here so that the present paper will be self-contained.
The full wave function $\mid \pi\rangle$ is dominated by components in
which the separation between the constituents is of the order of the
diameter of the physical pion. But there is a perturbative tail in
momentum space which accounts for the short distance part of the
pion wave function. This tail is of dominant importance
here because  we take the overlap with a  final state  constructed
from constituents moving at high relative momentum. It is therefore
reasonable to start our considerations from the one gluon exchange
contribution $V^g$, and use the light cone gauge:
$p'_{\pi~\mu}A^\mu=0$ and the Brodsky-Lepage \cite{BL} normalization
and phase-space conventions. Here  $p'_{\pi}=p_{\pi}-c\cdot p_{N}$
where $c$ is determined from the condition: ${p'_{\pi}}^2 =0$.
Evidently for  sufficiently high energies of pion projectile such a
gauge will almost coincide with the gauge $A_{-}=0$ and therefore  the
Wilson line operator between $q\bar q$ pair:
$\exp (i\int A_{\mu} dx_{\mu}) \approx exp (i\int A_{t} dx_{t})$.
Thus in the target rest frame, at the 
light cone where $x_{-}=0$ the Wilson line
operator does not produce additional gluons in spite of the large and
increasing with energy longitudinal distances  $x_{+}$. The wave
function $\chi(k_t,x)$ is gauge invariant because it depends on
transverse, i.e. on physical degrees of freedom. (On the contrary in the
Feynman gauge  the $q\bar q$ pair evolves into a many particle  state because
of the Wilson line operator. So a calculation in the
Feynman gauge should include the evaluation of the form factor which
guarantees dominance of the two particle final state.)
Another advantage of the gauge chosen in the paper is that the
amplitude for hard gluon exchange has no infrared divergences and
therefore  the separation between large and small distances is
straightforward. (On the contrary in the gauge $A^+=0$ the amplitude
for hard gluon exchange is infrared divergent.  The divergent contribution is,
however, exactly canceled out with that  in the fermion propagator in
this gauge, cf.\cite{BL}.) The chosen gauge is convenient to evaluate
the high
momentum  component of the pion wave function. For the evaluation of the
parton distribution within a nucleon another gauge is more
appropriate. This mismatch does not introduce, however, additional problems
because the factorization theorem justifies the 
possibility to choose independently
the gauges for the amplitude of pion fragmentation into jets and for the
parton distribution within a nucleon.
The perturbative tail is obtained as the result of the one gluon exchange
interaction acting on the soft  part of the momentum space wave
function, defined as
\bea
\psi(l_t,y)\equiv \langle l_t, y\mid \pi\rangle_{q\bar q}.
\label{psidef}\eea
By definition, $\psi$ is dominated by its non-perturbative
low-momentum components. However, the amplitude we compute depends
on the high momentum tail, $\chi$. For this component, perturbation theory is
applicable and we use the one-gluon exchange approximation to the
exact $q\bar{q}$ wave function of Eq.~(\ref{psidef}) to obtain
$\chi$, valid for large values of $\kappa_t$, as
\begin{eqnarray}
\chi.(k_t,x)={-4\pi C_F}
{1\over \left[m_\pi^2-{ k_t^2 +m_q^2\over x(1-x)}\right]}
\int_0^1 dy\int {d^2 l_t\over2(2\pi)^3} V^g(k_t,x;l_t,y)
\psi(l_t,y)
\end{eqnarray}
with
\begin{eqnarray}
V^g(k_t,x;l_t,y)=\alpha_s {\bar u (x,k_t)\over\sqrt{x}}
\gamma_\mu {u(y,l_t)\over \sqrt{y}}
{\bar v (1-x,-k_t)\over\sqrt{1-x}}\gamma_\nu
{v(1-y,-l_t)\over  \sqrt{1-y}}d^{\mu\nu}
\nonumber\\
\times \left[ {\theta(y-x)\over y-x}
{1\over m_\pi^2-{k_t^2+m_q^2\over x}-{l_t^2+m_q^2\over 1-y}-
 {(k_t-l_t)^2 \over y-x}}+(x\to 1-x, y\to 1-y)
 \right], \label{vgbl}
      \end{eqnarray}
where $C_F={n_c^2-1\over 2n_c}={4\over 3}$,  and $d^{\mu\nu}$ is the
projection operator of the gluon propagator evaluated in the light
cone cone gauge defined above :
$d_{\mu,\nu}=\delta_{\mu,\nu}-\frac{p'_{\mu}(\pi) k_{\nu}
+k_{\mu} p'_{\nu}(\pi)}{(p'_{\pi}k)}$.
Here $k$ is the gluon four-momentum.
The range of integration over  $l_t$ is restricted by the
non-perturbative pion wave function $\psi$.

Then in the evaluation of $V^g$  we set $m_q$ and $l_t$ to 0 everywhere
in the spinors and energy denominators. This is legitimate because of
a lack of infrared divergences in the amplitude of hard process 
\cite{hallertext}.

Thus
\begin{eqnarray}
V^g(k_t,x;l_t,y)\approx
{\alpha_s(k_t^2)\over x(1-x) y(1-y)}V(x,y)
\end{eqnarray}
where in the lowest order over coupling constant $\alpha_s$ but
keeping leading power of $k_t^2$:
\bea
V(x,y)=2\left[\left(\theta(y-x)x(1-y)+\theta(y-x) \right)+(x\to 1-x, y\to 1-y)
\right],\eea
We want to draw attention that in the gauge chosen in the paper
dominant contribution arises from the components of gluon propagator
transverse to the directions of pion and nucleon four momenta.

The net result for the high $k_t$ component
of the pion wave function is then
\bea
\chi(k_t,x)={4\pi\;2\; C_F \alpha_s(k_t^2)\over k_t^2}
\int_0^1 dy V(x,y){\phi(y,k_t^2)\over y(1-y)}
\label{pieq1},\eea
where nonperturbative support
\begin{equation}
\phi(y,k_t^2)\equiv
\int {d^2 l_t\over2(2\pi)^3} \theta(k_t^2-l_t^2)\psi(l_t,y).
\label{ph1}
\end{equation}

The analysis of experimental data for virtual Compton scattering
and the pion form factor  performed in \cite{tolya,kroll}
shows that this amplitude  is not far from the asymptotic one
\cite{arrr} for $k^2_t\geq 2-3$  GeV$^2$,
\bea
\phi(k_t^2\to\infty, y)=a_0y(1-y),
\label{phi}\eea
where $a_0=\sqrt{3}f_\pi$ with $f_\pi\approx 93$ MeV.

 Equation~(\ref{pieq1}) represents the high relative momentum  part of the
pion wave function in the lowest order of pQCD when running of
the coupling constant is neglected. Using the asymptotic function (\ref{phi}) in
Eq.~(\ref{pieq1}) leads to an expression for $\chi(k_t,x)\propto x(1-x)/k^2_t$
\bea
\chi(k_t,x)={4\pi C_F \alpha_s(k_t^2)\over k_t^2}a_0x(1-x)
\label{wave}.
\eea

The next step is  to use  the renormalization invariance of theory
to include the  $k_t$ dependence of the coupling constant :
\bea
\alpha_s(k_t^2)={4\pi\over \beta \ln{k_t^2\over \Lambda^2}},
\label{pieq11}
\eea
with $\beta=11-{2\over 3}n_f$. This can be easily done similar to
\cite{BL} where the relationship between $\phi(Q^2,x)$ and $\chi(k_t,x)$
and QCD evolution equation for $\phi(Q^2,x)$ have been deduced.

The quark distribution function $\phi(x,Q^2)$ -the amplitude for finding
constituents with longitudinal momentum $x$  in the pion which are
collinear up to the scale $Q^2$ is \cite{BL}
\bea
\phi(x,Q^2)={1\over d_f(\kappa_t^2)}
\int_0^{Q^2}\chi(k_t^2,x){dk_t^2\over 16\pi^2}.
\label{RELATION}
\eea
The factor :
${1\over d_f(\kappa_t^2)}=(\ln{Q^2\over \Lambda_{QCD}^2})^{-\gamma_F/\beta}$
arises from vertex and self-energy corrections. (By definition
the running coupling constant includes  the renormalization factor of
the  fermion
propagator. Such a renormalization factor is absent in the definition
of $\chi(k_t,x)$). One of advantages of the gauge chosen in this paper
is that $\gamma_F$
\begin{eqnarray}
\gamma_F=C_F,
\end{eqnarray}
has no infrared divergences in difference from \cite{BL} where gauge
$A_{+}$ has been chosen. Calculations are significantly
simplified
because the dominant contribution is given by transverse gluon
polarizations only. The 
evolution equation for $\phi(x,Q^2)$ gives \cite{BL,arrr}
\bea
\phi(x,Q)=x(1-x)\sum_{n=0}^{\infty} a_n C_n(1-2x)
\left({\ln Q^2\over \Lambda_{QCD}^2}\right)^{-\gamma_n},
\eea
where $\gamma_n={C_F\over \beta}
(1 -{2\over (n+1)(n+2)}+4\sum_{k=_2}^{k=n+1} 1/k )$, $C_n(1-2x)$ are Gegenbauer
polynomials . Coefficients $a_n$ are the subject of discussions in the
literature and they can be estimated within the current models. The above
equation makes it possible to calculate the
high $k_t$ behavior of $\chi(k_t,x)$ as
\bea
{\chi(k_t^2=Q^2,x)\over 16 \pi^2}=
{d \left[\phi(x,Q)\left(ln{Q^2\over \Lambda_{QCD}^2}\right)^
{\gamma_F/\beta}\right]
\over d{Q^2}}.
\label{asymptotics}
\eea
Thus the  $z,\kappa_t$ dependence of the cross section for the diffractive
dijet production should be sensitive at moderately large $k_t$ to the terms
involving  Gegenbauer polynomials of the order greater than 0 in
the pion wave function if $a_n$ are sufficiently large. The process of
the photoproduction of jets at HERA will be appropriate for this purpose.

It follows from the above equation that the 
asymptotic wave function is as follows:
\bea
\chi(k_t,x)={4\pi C_F \alpha_s(k_t^2)\over k_t^2}a_0x(1-x)
\left(ln{Q^2\over \Lambda_{QCD}^2}\right)^{C_F/\beta}
\label{asymptoticwf}.
\eea
Thus QCD predicts the dependence of $\chi(k_t,x)$ to be of the form
used in Ref.~\cite{fms93}.

\subsection{Interaction with the target}
To  compute the amplitude $T_1$ it is necessary to  specify the
scattering operator $\widehat f.$ We will fix the transverse
recoil  momentum of the target at zero to  simplify the discussion.
The transverse distance operator
$\vec b =(\vec{b}_{q}-\vec{b}_{\bar{q}})$ is canonically conjugate
to $\vec{\kappa}_t$. At sufficiently small values of $b$,  the
leading twist effect
and  the dominant term at large $s$ arises from  diagrams in which the
pion fragments into two  jets as a result of interactions with
the two-gluon component of the gluon field of a target, see Fig. 1.
The perturbative QCD determination of this interaction, which is a type
of  QCD factorization theorem, involves a diagram similar to the gluon
fusion contribution to the nucleon sea-quark content observed in deep
inelastic scattering. One calculates the box
diagram for large values of $\kappa_t$ using the wave function
of the pion instead of the vertex for $\gamma^*\to q\bar q$.
The application of the technology leading to the
QCD factorization theorem in the impact parameter space
leads \cite{fmsrev,fms93,bbfs93,frs} to
\begin{equation}
   \widehat f(b^2)=i s \frac{\pi^2}{3} b^2 \left[ x_N
 G_N(x_N, Q^2_{\rm eff}) +2/3 x_N S_N(x_N, Q^2_{\rm eff}
 ) \right] \alpha_{s}(Q^2_{\rm eff}),
\label{eq:1.27}
 \end{equation}
in which
$G_N$ is the gluon distribution function of the nucleon,
$S_N$ is the sea quark distribution function of the nucleus for a
 flavor
coinciding with that of the $q\bar q$ dipole, and
$Q^2_{\rm eff}={\lambda\over b^2}$. The factor 2/3 appearing in the
second term is the same as in the LO approximation
for the longitudinal structure function and exclusive vector
meson production \cite{Lech}. The only difference is
that in our case the number of flavors is unity.
For our kinematics, it is reasonable
to use $\lambda(x=10^{-3})=9$  \cite{fks}. The formula (\ref{eq:1.27})
should be modified when applied to  hard diffractive processes. The
mass difference between the pion and the final two-jet state requires that
the reaction proceeds by  a non-zero momentum transfer to the target.
This means that the function $G_N$ should be replaced by the skewed
(or off-diagonal or generalized) gluon distribution. Thus the
distribution function  should depend on the plus components $x_1,x_2$
of the momenta $k_1,k_2$ of the two exchanged gluons, Eq.~(\ref{x1x2}).

The difference between the skewed and ordinary gluon  distribution
is calculable in QCD using the  evolution equation
for the skewed parton distributions\cite{FFGS,fg}.
The kinematical relation between $x_1$ and $x_1-x_2$ is given in
Eq.~(\ref{diff1}). But $x_1$ is close to $x_N$ of Eq.~(\ref{xn}),
while $x_2$ is small  in the calculation of $T_1$.
The skewed parton distribution can be approximated
by a gluon distribution \cite{abramowicz,radyushkin} if
\bea
x_N\approx (x_1+x_2)/2\approx
{\kappa_t^2\over 2 z(1-z)s}
.\label{xn}
\eea
While including the  effect of skewedness would alter  any
detailed numerical results, the qualitative features of
the present analysis would not be changed.

The most important effect shown in Eq.~(\ref{eq:1.27}) is the $b^2$
dependence, which shows the diminishing strength of the
interaction for small values of $b$. To simplify formulas
it is convenient to rewrite $\sigma$ in the form:
\bea
 \widehat {f}(b^2)=is {\sigma_0\over
\langle b_0^2 \rangle} {b^2 }
 =is {\sigma_0\over \langle b_0^2 \rangle}\;
\left({-\nabla^2_\kappa}\right), \label{fb2}
 \eea
in which the logarithmic dependence of $\alpha_s$ and the gluon distribution
on $b^2$ are included in $\sigma_0$. It is easy to check by direct
calculations of  Feynman diagrams that the operator $\nabla^2_\kappa$ acts
on the transverse momentum variables of the  pion wave function.
Our notation is that $ \langle b_0^2 \rangle$ represents the pionic
average of the square of the transverse separation, and
within the leading log accuracy
\bea{\sigma_0 \over \langle b_0^2 \rangle}
\approx{\pi^2\over 3}
\alpha_s(\kappa_t^2)
[x_N\;G_N^{(skewed)}(x_1,x_2,\kappa_t^2)],
\label{ord}\eea
in which the ordinary gluon distribution is used as the  initial
condition for the QCD evolution equation for the skewed/generalized
gluon density.  This result is a factor of four smaller than presented
in Ref.\cite{fms93}.

The  result (\ref{ord}) holds for $x_N$ about $10^{-2}$. For $x_N$
of  about $10^{-3}$ or smaller, the second kinematic
regime mentioned in the introduction is relevant, and   one
would obtain different results. For still smaller values of $x$, say
$x\sim 10^{-5}$ non-linear gluonic effects become important, and the
present treatment of the $q\bar q$ interaction with the target
may be  insufficient.

\subsection{One Gluon Exchange in the Pion-- $T_1$}
The necessary inputs to evaluating $T_1$  are now
available. The approximate pion wave function, valid for large
relative momenta, is given by  Eq.~(\ref{wave}). The interaction
$\widehat {f}$ is given by Eq.~(\ref{fb2}). The use of Eq.~(\ref{fb2})
allows a simple evaluation of the scattering amplitude $T_1$ because
the $b^2$ operator acts on the pion wave function (here $\sigma_0$ is
treated as a constant) as  $-\nabla_{\kappa_t}^2$,  leading to the result
 \bea
T_1=-4is{\sigma_0\over \langle b^2 \rangle}{4\pi C_F\alpha_s(\kappa_t^2)
   \over \kappa_t^4}
(\ln{k_t^2\over \Lambda_{QCD}^2})^{C_F\over \beta}\;a_0\;z(1-z).
\label{t1r} \eea
This amplitude $T_1$ is of the same form as our 1993 result\cite{fms93}.
The present result is obtained directly from QCD, in contrast with
the earlier work which used some phenomenology for the pion wave function.

Corrections to Eq.(\ref{t1r}) are of the order
${1\over \ln{\kappa_t^2\over \Lambda^2}}$.
For example, a literal application of
Eq.~(\ref{fb2})  would lead to a factor
$\left(1+{2\over \ln{\kappa_t^2\over \Lambda^2}}\right)$. However,
similar corrections may arise from other effects not considered here.
So a calculation of such corrections is beyond the scope of this paper.

Our $\kappa_t$  dependence of $T_1$ leads to:
${d\sigma(\kappa_t)\over d\kappa_t^2}\propto
\frac{\left(\ln ({k_t^2\over \Lambda_{QCD}^2})\right)^{2C_F/beta}}{\kappa_t^8}$
for $x_N\sim 10^{-2}$. This can be understood using
simple reasoning. The probability of finding  a pion at
$b\leq {1\over \kappa_t}$ is $ \propto b^2$, while the square of the
total cross section for  small-dipole-nucleon interactions
is $\propto b^4$. Hence the cross section of productions of jets with
sufficiently large values of $\kappa_t $
integrated over $d^2 k_t$
is $\propto {\alpha_s(k_t^2)^2
\left(\ln ({k_t^2\over \Lambda_{QCD}^2}\right)^{2C_F\over \beta}
\over \kappa_t^6}$,
for $x_N\sim 10^{-2}$, leading to a differential
cross section $\propto {1\over \kappa_t{^8}}$. This reasoning ignores
the dependence of the gluon structure function on $\kappa_t$. For
sufficiently small values of $x (x\sim 10^{-3})$,
gluon evolution would give a somewhat different behavior.

\section{Nuclear Dependence of the  Amplitude}

The picture we have obtained  is that the
amplitude is dominated by a process
in which the pion becomes a $q \bar{q}$ pair of
essentially zero transverse extent well
before hitting the nuclear target.
This point-like configuration (PLC) can
move through the entire nucleus without
expanding. The $q \bar{q}$ can interact
with one nucleon and can pass
undisturbed through any other nucleon. For
zero momentum transfer $q_t$
to the nucleus, the amplitude ${\cal M}$ (A) takes the form
\begin{equation}
{\cal M} {\rm (A)} = {\rm A}\; {\cal M} (N)
{G_A(x_1,x_2,m_f^2)\over AG_N(x_1,x_2,m_f^2)}
\left(1 + \frac{\epsilon}{<b^2>
\kappa_t^2} {\rm A}^{1/3}\right)
\equiv {\rm A} {\cal M} (N) \Gamma \; ,
\label{final}\end{equation}
in which the
skewedness of the gluonic
distribution is made explicit, and
where the real number $\epsilon > 0$. Observe the
factor $A$ which is the dominant effect here.
This factor is contained in
the ratio of gluon distributions in a nucleus
and in a nucleon\cite{fms93}.
This dependence on the atomic number is a reliable
prediction of QCD in the limit
$m_f^2$ and $s\to\infty$, with
fixed $x_N$ (of Eq.~(\ref{xn})). (The quantity $m_f$ is the mass of the dijet
system.) On the contrary, for
$x_{N}\to 0$ with  fixed $m_f^2$, the
nuclear  shadowing of the gluon distribution
becomes very important
\cite{fms93,bfgms}.

The $\epsilon$ correction term in Eq.~(\ref{final})
is a  higher twist contribution which arises from a single rescattering
which can occur as the PLC moves through the nuclear
length $(R_A \propto$ A$^{1/3}$).  That $\epsilon>0$
was a somewhat surprising
feature of our 1993 calculation because the usual second order
rescattering, as treated in the Glauber theory, always  reduces cross sections.
This highly unusual sign follows
from the feature in QCD that the  relative contribution of the rescattering
term (screening term)
 decreases with increasing size of
the spatially small dipole.
The key features of the usual first order term are
$f = i \sigma$ , and those of
the usual second order term are $if^2 = -i\sigma^2$.
Note the opposite signs. For us here $f = i \sigma_0 b^2/<b^2>$,
For very large values of $\kappa_t^2$
the operator $b^2$, as applied to the pion wave function $\chi$, which falls
with
 $\kappa_t^2$ as $1/\kappa_t^2$ gives
$b^2 \chi= -\frac{4\chi}{\kappa_t^2}$. So the  first
term
$f\chi = -i 4 \sigma_0  /(<b^2> \kappa_t^2) \chi $  now has  the same sign as
the second--order term:
$if^2\chi = -i [\sigma_0/<b^2> b^2]^2\chi =
-i{32 [\sigma_0/(<b^2>\kappa_t)]^2\chi}$

The differential cross section takes the form
\begin{equation}
\frac{d\sigma(A)}{dq^2_t} = A^2\Gamma^2\frac{d\sigma(N)}{dt}
 e^{tR^2_A /3} \; ,\label{dsig}\end{equation}
for small values  of $t$. Note that
\bea
-t=q_t^2-t_{\rm min},
\eea
where $-t_{\rm min}$ is the minimum value of
the square of the longitudinal momentum transfer:
\bea
-t_{\rm min}=\left({m_f^2-m_\pi^2\over 2p_\pi}\right)^2.\eea
Our discussion below is applicable in the kinematics where
$-t_{min}R^2_A /3\leq 1$ so that the entire  dependence of the
cross section on $t_{min}$
is contained in the factor $e^{t_{min}R^2_A /3}$.

   One measures the integral
\begin{equation}
\sigma (A) = \int dt
\frac{d\sigma (A)}{dt} =
\frac{3}{R^2_A} A^2 \Gamma^2 \sigma (N) \; .
\end{equation}
A typical procedure is to parametrize $\sigma(A)$ as
\begin{equation}
\sigma (A) = \sigma_1 A^\alpha\label{param}
\end{equation}
in which $\sigma_1$ is a constant independent of A. For the
$R_A$ corresponding to the two
targets Pt (A = 195) and C(A = 12) of E791,
one finds $\alpha \approx 1.45$. The experiment\cite{danny} does not directly
measure the coherent nuclear scattering cross section. This must be extracted
from a measurement which  includes the effects of nuclear
excitation. The extraction is discussed below.

As pointed out previously\cite{hallertext}, the values of our multiple
scattering correction of our 1993 calculation\cite{fms93} were
overestimated by a factor of approximately four. This is because the
$\sigma_0$ was chosen to be larger by a factor of 4 than  in \cite{bbfs93}
and in  Eq.~(\ref{eq:1.27}). We now find that for values of
$\kappa_t$ greater than about 2 GeV, the coefficient $\alpha$
could be enhanced by between 0.0 and 0.08, depending on the value of
$\kappa_t$. Taking 0.04 as a mean value one finds $\alpha\approx 1.5$. This
estimate depends on the use of a model for the non-perturbative part of
$\vert\pi\rangle$, and also on the
validity of a simple eikonal treatment for the multiple-scattering
corrections which is questionable at the high energies that we consider here.

Another potentially important A-dependent effect is 
the nuclear shadowing of
the parton densities. Very little direct experimental information is available
for the A-dependence of the gluon structure function.  The analyses of the data
combined with the momentum sum rules and the calculation of gluon shadowing at
small $x$ suggest that   the value
$x\sim 0.01$ (which corresponds to the kinematics of \cite{danny})
is in a transition region between
the regime of an enhancement of the gluon distribution
at  $x\sim .1$ and the strong
shadowing at $x\leq 0.005$ \cite{FLS90,Pirner,Eskola,fs99}.
Using the A-dependence of $F_{2A}$ as a guide, and in particular the NMC
ratio $F_{2Sn}/F_{2C}$ \cite{NMC}, the shadowing may reduce
$\alpha$ for the \cite{danny} kinematics by $\Delta \alpha \sim -0.08$.

\section{Experimental Considerations}
The requirements for observing the influence of color transparency
were discussed in 1993\cite{fms93}. The two jets should have total transverse
momentum to be very small. The relative transverse momentum should be
greater than about $\geq$ 2 GeV and the mass of the  diffractive state should
be described by the formula
\begin{equation}
m_f^2 =\frac{m_{q}^2+\kappa_{t}^2}{z(1-z)}.
\end{equation}
Maintaining this condition is necessary  to 
suppress diffraction into a $q\bar q g$ pair in which
the gluon transverse momentum is not too small.

It would be nice if one could measure the jet
momenta precisely enough so as to
be able to identify the final nucleus
as the target ground state.
  While it is very feasible to consider
for eA colliders,  this is impossible for high energy
fixed target experiments, so another technique must be used here.
The technique  used in \cite{danny} is to isolate the dependence
of the elastic diffractive peak on the momentum transfer to the target, $t$,
as the  distinctive property of the coherent processes. This was done
by first introducing a cut on the momentum of the observed pions by requiring
that they carry more than 90\% of the incident momentum.
This sample was than analyzed as a function of the total transverse momentum
of the system. A strong coherent peak was observed with the slope consistent
with the coherent contribution. The background was fitted as a sum of
the coherent peak, inelastic diffraction with a nuclear break up and
the term due to inelastic events where some hadrons were not detected.

The amplitude for the
non-spin flip
 excitations of low-lying
even-parity nuclear levels
$\sim{-t}$, due to the orthogonality of the ground and
excited state nuclear wave functions. Thus the
cross section of
these kinds of soft nuclear excitations
integrated over $t$ is
  suppressed by an additional factor of
$1/R_A^4\approx A^{-4/3}$ compared to the
nuclear coherent process.
For $\sqrt{-t}\;R_A\gg1$ where $q$ is the
momentum transfer to the nucleus
$q=\sqrt{-t}$ the background processes
involving nuclear excitations
vary as A, so an unwanted counting of such would actually
weaken the signal we seek.
For $qR_A\gg1$ the diffractive peak cannot be
used as signature of
diffractive processes to distinguish them
from non-diffractive
processes whose cross section
$\propto \sigma(\pi N) \propto A^{0.75}$.
Thus substantial A-dependence, $\sigma \propto
A^{\alpha}$ (with $\alpha \approx 1.5$), as
predicted by QCD for large enough values of $\kappa_t$,
should be distinguishable from the background processes.

The amplitude varies as
${\cal M}(A) \sim \alpha_s/\kappa_t^4$
 \begin{equation}
{\cal M}(A) \sim \alpha_s x_NG_N(x_N,Q_{eff}^2)/\kappa_t^4,
\end{equation}
where $Q^2_{eff} \sim 2\kappa_t^2$. For the
kinematics of  the E791 experiment,
where $x_N$ increases $\propto \kappa_t^2$, the
factor $\alpha_s x_NG_N(x_N,Q_{eff}^2)$ is a rather
weak function of $\kappa_t$. For example,
 if we use the standard CTEQ5M parameterization we find
 $\sigma (A) \sim 1/ \kappa_t^{8.5}$  for $1.5 \leq \kappa_t
\leq 2.5 $ GeV which is consistent with
the data
\cite{danny}.
For the amplitude discussed here,
$\sigma (A) \sim ( z(1-z))^2$ which is in the excellent agreement
with  the data
\cite{danny}.

\subsection{Extracting the coherent contribution}

The experiment is discussed in Ref.~
\cite{danny}.
The main advantage of
this experiment  is  the
excellent resolution of the transverse
momentum. The reference also
shows the identification of the dijet using
the Jade algorithm, and it
displays the identification
of the diffractive peak by the $q_t^2$ dependence for very low $q^2_t$. This
dependence is consistent with that obtained
from the previously measured radii
$R(C) = 2.44$ fm, and $R_{Pt} = 5.27$ fm. The key
feature is the identification
of the coherent contribution from its rapid
falloff with $t$.

 \begin{figure}
\begin{center}
\epsfig{file=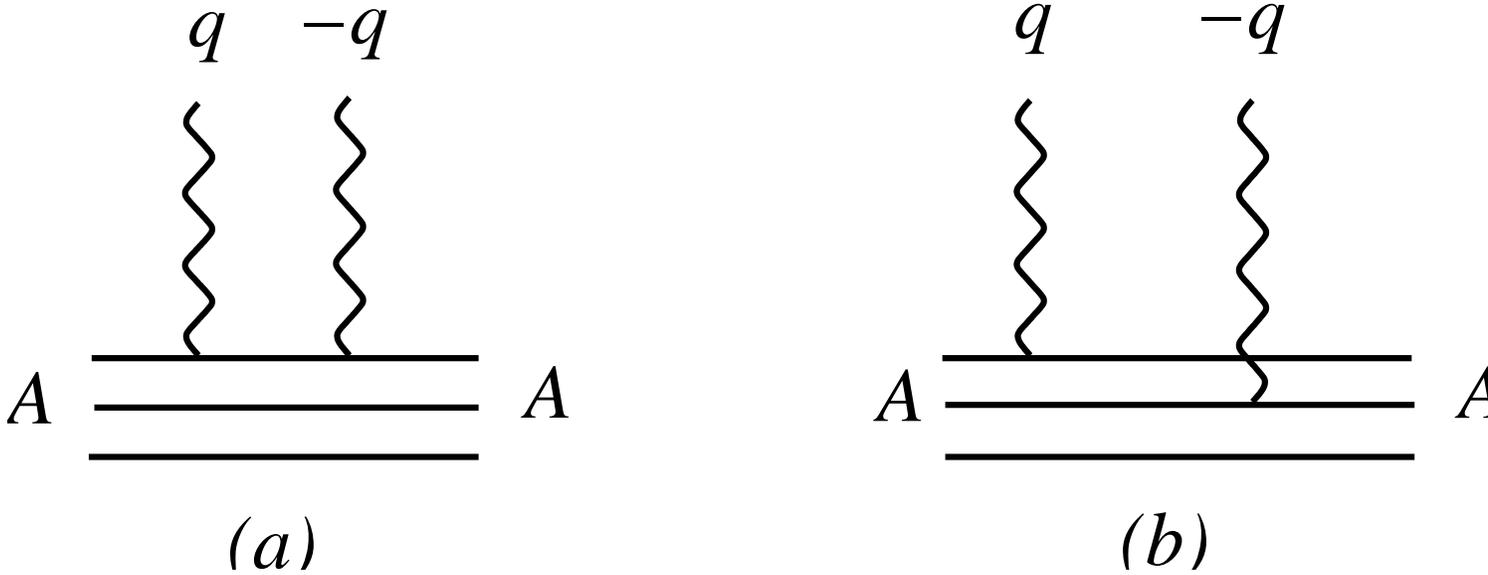, height=3.0in}
\vspace{0.3in}
\caption    { Contributions to the total nuclear
        diffractive cross section.
The wavy lines in  this figure denote amplitude for the scattering of $q\bar q$
pair by  a nucleon.
(a) The terms with $i=j$. (b) $i\neq j$.}
     \end{center}
\end{figure}

We discuss the extraction of  this  signal in some detail.
We consider the contribution to the total
cross section ${d\sigma_A\over dt}$
that arises from the diffractive production of the dijet.
The total cross section includes terms in
which the final nucleus is not the ground state. The nuclear
excitation energy is small
compared to the energy of the
beam, so that one may use closure to treat the
sum over nuclear excited states.
Then the cross section is evaluated as a ground
state matrix element of an
operator
$\sum_{i,j}e^{i\mbox{q}\cdot(\mbox{r}_i-\mbox{r}_j)}=A+
  \sum_{i\neq j}e^{i\mbox{q}\cdot(\mbox{r}_i-\mbox{r}_j)} \;$;
see Fig.~13. The result, obtained by using $\mbox{r}_i$ relative to
the nuclear center of mass, and by
neglecting correlations in the nuclear wave function is given by
\bea
{d\sigma_A\over dt}=\left[A +A(A-1)F_A^2\left(t\; {A\over A-1}\right)\right]
{d\sigma_N\over dt}.
\eea
The factor ${A\over A-1} $ in the argument of $F_A$
is due to
accounting
for nuclear
recoil in the mean field approximation, {\em cf.} \cite{BW}.
This formula should be very accurate, for the
small values of $t$ relevant here. The contribution of
the coherent processes to the
total cross section is given by
\bea
{d\sigma^{coherent} \over dt}=A^2\;F_A^2(t){d\sigma_N\over dt},
\eea
and the contribution of excited
nuclear states is the difference:
${d\sigma_A\over dt}-{d\sigma^{coherent}\over dt}$, which vanishes at $t=0$.
The experiment proceeds by removing a
term $\propto A$ from ${d\sigma_A\over
  dt}$ which has no rapid
variation with $t$. This defines  a new cross section
which is actually measured experimentally.
\bea {d\tilde{\sigma}_A\over
  dt}=A(A-1)F_A^2\left(t\; {A\over A-1}\right){d\sigma_N\over dt}.
\eea
The integral of this term  over $t$
can be extracted from the data:
\bea \sigma_1\equiv \int\;dt\; {d\tilde{\sigma}_A\over dt}
={3A(A-1)\over r_N^2 +R_A^2{A\over A-1}}
{d\sigma_N\over dt}_{\left|t=0 \right.}
\approx {3(A-1)^2\over r_N^2 +R_A^2} {d\sigma_N\over dt}_{\left|t=0 \right.}.
\label{late}\eea
Here the factor $r_N^2$ takes into account the slope of
the elementary cross section,
assuming that it is determined solely by the nucleon vertex.
Note that the result (\ref{late})
differs by a factor of ${(A-1)^2\over A^2}$ from the
$A$-dependence predicted  previously
for coherent processes,
 recall Eq.~(\ref{dsig}). Using A=12, 195, the nuclear radii mentioned above,
$r_N=0.83 fm$ and fitting the ratio of cross sections obtained from
Eq.(\ref{late}) with
the parametrization $\sigma\propto A^\alpha$, gives then
\bea \alpha=
1.54,
\label{ctr}\eea instead of $\alpha=1.45$.

The result \cite{danny} of the experiment is
\begin{equation}
\alpha \approx 1.55 \pm 0.05 \; ,
\end{equation}
which is remarkably close to the theoretical value shown in
Eq.~(\ref{ctr}).
The sizes of our multiple scattering and nuclear shadowing
corrections, which work in the opposite directions,
and which were
discussed in the previous Section, are of the order of
the experimental error bar.

\section{Electromagnetic Background}

Because very low values of $q^2_t$ are involved, one could ask if the process
occurs by a one-photon exchange (a type of Primakoff effect) instead of
a two-gluon exchange. If the momentum transfer is very low,
the process is peripheral and there would be no initial or
final state  interactions. Thus, it is necessary
to estimate the  relative importance of the two effects.

This amplitude is caused by the exchange of a
virtual photon of four-momentum $q$ ($q^2=t$) with the target. The
nuclear  Primakoff amplitude is then given by
\begin{equation}
{\cal M}_P (A) = e^2 {\langle\pi\vert J_\mu^{\rm em}
\vert q\bar{q}\rangle\over -t}
  (P_A^i+P_A^f)^\mu {Z\over A}F_A(t)\approx {2 e^2}
 \langle\pi\vert J^{\rm em}\cdot {P_A\over A}
     \vert q\bar{q}\rangle {ZF_A(t)\over -t},\label{prim}
\end{equation}
where we will use the decomposition $-t=q^2=q_t^2-t_{\rm min}$.
A photon can be attached to any charged particle, so a direct calculation
would involve a complicated sum of diagrams. We
may simplify the calculation by using the fact that the
electromagnetic  current is conserved. This
application is simplified by the use of Sudakov variables, which is a
necessary first step. Accordingly, we write
\bea
q=\alpha{P_A\over A}+\beta p_\pi+q_t.
\eea
Conservation of four-momentum gives
\bea \beta={q^2\over 2(p_\pi\cdot P_A)},\quad\alpha
={-m_f^2\over s}.\label{kin}\eea
Then conservation of current can be written as

\bea  \langle\pi\vert  J^{\rm em}
\cdot q\vert q\bar{q}\rangle\approx
 \alpha \langle\pi\vert  J^{\rm em}
\cdot{P_A\over A}\vert q\bar{q}\rangle
+\beta \langle\pi\vert  J^{\rm em}\cdot
p_\pi\vert q\bar{q}\rangle-\langle\pi\vert
J^{\rm em}\cdot q_t\vert
q\bar{q}\rangle=0.
\label{cc}\eea
The use of Eq.~(\ref{kin})
and keeping only the leading term in   $\mu^2/s$,
  where $\mu$ is the typical mass
involved in the considered process,
allows us to neglect
the $\beta$ term of
Eq.~(\ref{cc})
so that
\bea \alpha
\langle\pi\vert  J^{\rm em}\cdot
{P_A\over A}\vert q\bar{q}\rangle=
\langle\pi\vert  J^{\rm em}\cdot q_t\vert
q\bar{q}\rangle
\label{wf}\eea
By definition, the  transverse momentum of a pion is zero,
so the dominant (in terms of powers of $\kappa_t$)
contribution in  Eq.~(\ref{wf}) is given by photon
attachments to quark lines, and  the matrix element is given by
\footnote{The above  (\ref{above}) differs from that of
the first version of our  paper which appeared in hep-ph. We are indebted to
D.Ivanov and L.Sczymanowsky
- who drew our attention to the  misprint in  this version of the paper.}
\bea
\langle\pi\vert  J^{\rm em}\cdot q_t\vert
q\bar{q}\rangle=\chi_\pi(z,\kappa_t)q_t\cdot\kappa_t (2/3z-1/3(1-z))
\label{above}.\eea
The generalization of this result to account for all Feynman diagrams
having the same  powers of s and $\kappa_{t}^2$ is almost trivial.
The  relative contribution of other diagrams is $\propto k_{t}'/\kappa_{t}$
where $k_{t}'$ is the transverse momentum of quarks in the intermediate state.
But within the $\alpha_{s} \ln \kappa_{t}^2/\Lambda_{QCD}^2$ approximation
$k_{t}'^2\ll \kappa_{t}^2$ so this contribution does not lead to
a $\ln \kappa_{t}^2/\Lambda_{QCD}^2$ term. Thus the  above formula is valid
within the $\alpha_{s} \ln \kappa_{t}^2/\Lambda_{QCD}^2$ approximation
when $\alpha_{s}\ll 1$.

The net result, obtained by using Eq.~(\ref{wf})
in  Eq.~(\ref{prim}),  is
\begin{equation}
{\cal M}_P (A) = \frac{-e^2 {\chi}_\pi
(z,\kappa_t) Z}{q^2_t-t_{\rm min}}F_A(t){s\over k_t^2}
2q_t\cdot\kappa_t (2/3-z)\;
\end{equation}
which should be compared with the amplitude of
Eq.~(\ref{t1r}) (including the
effect of the nuclear form factor, which enters
at non-zero values of $t$, but
ignoring the logarithmic correction) written as
\begin{equation}
{\cal M} (A) = -i {\chi}_\pi (z,\kappa_t) A
\frac{s\sigma_0 \nabla^2_{k_t}}{<b^2>}F_A(t)
\approx  4 i {\chi}_\pi (z,\kappa_t)/k_{t}^2 A
\frac{s\sigma_0 }{<b^2>}F_A(t)
\label{str}
\end{equation}

The Primakov term has been evaluated also in the paper of D.Ivanov and
L.Sczymanowsky \cite{IS1}. As the consequence of different approximations
result obtained in \cite{IS1} differs from ours. We use conservation of 
the e.m.
current to deduce the Weizecker-Williams approximation and to account for the
cancellation between diagrams corresponding to attachments of photon to the 
different charged constituents of a pion. After accounting for the
cancellation we restrict ourselves by the contribution of diagrams enhanced
by the large factor $\kappa_t$ from the vertex of quark(antiquark)- photon
interaction. Dominance of photon interaction with external quark lines
in the pion fragmentation into 2 jets is another
form of the QCD factorization theorem which properly accounts for 
the  conservation
of the e.m. current and the 
gauge  invariance of QCD. On the contrary \cite{IS1} 
interacting particles are put on mass shell before the 
separation of scales and
the 
cancellations between different photon attachments has been taken into account.
These approximations have problems with the conservation of the 
e.m.current and
the renormalization group in QCD.

The ratio of electromagnetic  and strong amplitudes is given by
\begin{equation}
\frac{{\cal M}_P (A)}{{\cal M} (A)} =
-i\;e^2\;{ Z\over A}\;
{{2/3-z}\over\sigma_0/ <b^2>}\;
{q\cdot\kappa_t\over 2(q_t^2-t_{\rm{min}})}
\end{equation}
Using $q^2_t \approx 0.02$ GeV$^2$ (the smallest
value measured in
\cite{danny}
Z/A=1/2,
$e^2 = 4\pi/137 , \kappa_t = 2$ GeV, $\sigma_0 / <b^2> \approx 2.5$
(this is 1/4 of the value of Ref.~\cite{fms93}),
and taking $q_t$ parallel to $\kappa_t$ we find that
\bea
\frac{{\cal M}_P (N)}{{\cal M} (N)}
\approx -1/8(2/3-z) \;i\approx -0.02 \;i
,\eea
with $z=1/2$.

Thus the Primakoff term is very small and, because of its real nature,
interference  with the larger strong amplitude (which is almost
purely imaginary) is additionally suppressed. We may safely ignore this
effect for the energy range of Ref.~\cite{danny}, or
any contemplated fixed target experiment.
At collider energies, it will be possible to measure jets of
much larger values of transverse momentum, so any complete
theoretical  analysis should account for this electromagnetic  interaction.

For a heavy nuclei target another electromagnetic process
$\propto Z^2\alpha_{em}^2$ in the amplitude gives a contribution.
This is a dijet production due to the two-photon 
exchange, a version of Fig.~8 in which the  exchanged gluons are
replaced  by exchanged  photons.  For small transverse momenta of
quarks $l^2_{t}\ll \kappa_{t}^2$ in the pion wave
function, this contribution is suppressed
as the power of $s$ in  the
amplitude. This is because, in the
calculation of the imaginary part of the 
diagram, $x_{2}$ for the exchanged photon is given by
\begin{equation}
x_{2}\nu=-{k_{t}^2\over z},
\end{equation}
except for a very narrow region of $z$ near $z=0,1$, which is
suppressed by the decrease of the 
pion wave function. Thus $x_2<0$ and according to our previous arguments, this
contribution should be very small. For large $l_{t}$
this contribution may  also be
expressed in terms of  the same pion wave function as in
the case of the two-gluon exchange. It is easy to estimate
this EM amplitude obtained from the two-photon exchange:
\begin{equation}
{\cal M} (\gamma\gamma)
= \alpha_{em}^2 (s/k_t^2)
{\chi(z,k_t)}_\pi  Z^2
\int {d^2 l_{t}\over l_t^2} F_A(l_t)F_A(q_t-l_t)
(8 e_1 e_2)(1-2z).
\end{equation}
Here $e_1(e_2)$ is electric charge of quark (anti-quark) in
the units of the electric charge of electron. The lower limit of integration
over $l_t^2$ is  $ (\kappa_t^2/2z(1-z))^2$
Thus the contribution of this term to the cross section
should have the same $z$ and  $\kappa_{t}^2$
dependence as the two-gluon exchange term, but
with a much faster Z dependence ($\propto Z^4$).
This term is negligible also.

\section{Discussion and Summary}

The use of the experimentally measured \cite{danny}
value of $\alpha = 1.55$ (recall    Eq.~(\ref{param})) leads to

\begin{equation}
\frac{\sigma (Pt)}{\sigma (C)} = 75 \; .
\end{equation}
The typical usual nuclear dependence of the soft diffractive processes
observed in the high energy processes is
$\approx A^{2/3}$ or $ \alpha =2/3$.
The use of a  Glauber approximation with a 
typical hadronic cross section for the
final system tends to predict the A dependence as $\approx
A^{1/3}$. An 
account of color fluctuations \cite{glauberadep} predicts $\approx A^{2/3}$,
in agreement with the FNAL data \cite{Ferbel} which  would give

\begin{equation}
\frac{\sigma_{\rm USUAL} (Pt)}{\sigma_{\rm USUAL} (C)} = 7.
\end{equation}
Thus color transparency causes a factor of 10 enhancement! This seems to be
the huge effect of color transparency that many of us have been hoping to
find. It is also true that, as noted in the Introduction, that
the $\kappa_t$ and $z$ dependence of
the cross section  \cite{danny} is in accord with our prediction.

All of this looks very good, but it is necessary to provide some words
of caution. Our analysis was related to a nuclear coherent
process involving a $q\bar q$ final state. If the experimental signal is
significantly contaminated by incoherent nuclear effects or by
$q\bar q g$ final states, our analysis might
not be applicable. However, the experimental \cite{danny}
extraction of the coherent peak using the $q_t^2$ dependence of the
amplitude, and the measurement of the two-jet (as opposed to three-jet)
cross section seem very secure to us, except for the
small correction discussed in Sect.~V. Another worry is that
the color transparency effect seen in Ref.~\cite{danny}
seems to start for values of $\kappa_t $ near 1 GeV.
These are lower than suggested in Ref.~\cite{fms93}. These earlier
predictions used modeling of non-perturbative effects, and such modeling
may be necessary to guess the lowest values
of $\kappa_t$ for which color transparency would occur.
The reasoning of the present paper uses perturbative QCD, which becomes
more reliable as $\kappa_t$ increases. This is because the competing
amplitudes $T_{2,3,4}$ are decreased relative to $T_1$ by a factor
of ${\Lambda^2\over \kappa_t^2}$  ($ \approx .04$ for $\kappa_t=1 $ GeV)
or $\alpha_s(\kappa_t^2)$. A coherent sum of
the sub-dominant amplitudes could provide a significant correction to
our dominant pure amplitude. However, the observed falloff of the
cross section with $\kappa_t$, combined with the $z$ and $A$ dependence, does
provide very strong evidence for color transparency.

It is worth noting similarities and differences between the process
we discuss in this paper  and another factorizable process, that  of hard
exclusive electroproduction of mesons \cite{bfgms,CFS}.
Both processes allow a simple geometric interpretation in the transverse
coordinate representation: a convolution of the initial wave short-distance
wave function, $\psi_{in}(z,b)$,  the dipole-target cross section,
$\sigma(b,x)$, and the final wave function $\psi_{fin}(b,z)$. However,
in the case of the vector meson production, the
$Q^2$ dependence
of the longitudinal photon wave function at small $b$ causes the final
vector meson wave function to be evaluated  at
$b\approx 2/Q$.
In the case of the pion dijet diffraction
the pion wave function enters at small values of
$b\sim {1\over \kappa_{t} }$.

If color transparency has been correctly observed in the 
$\pi + A \to q \bar{q} + A$ (ground state), there would be many implications.
The spectacular enhancement of the cross section would  be  a 
novel effect. The point-like configurations   would be proved to exist.
This would be one more verification of the concept and implications of
the idea of color. Furthermore, the definitive proof of the existence of
color transparency means  that we now have available a new effective tool
to investigate microscopic hadron, nuclear structure at hundreds of
GeV energy range.
At lower  energies of a ten's of GeV color transparency is masked
to some extent by the diffusion of a spatially small quark-gluon
wave package to the normal hadronic size \cite{fmsrev}. Still it
is possible that previous experiments \cite{[7]} showing hints
\cite{[8]} of color transparency (for a review
see Ref~\cite{fmsrev}) probably do show color transparency.
Efforts \cite{[9]}  to observe color transparency at Jefferson
Laboratory, and at HERMES(DESY) should be increased.
The electron-ion collider would  provide numerous possibilities for studying
color transparency both in di-, tri-jet coherent production
as well as in exclusive processes.

The observation of CT confirms  the idea that the life span of the
perturbative phase can be increased by the large Lorentz factor
associated with high energy beams. A challenging problem would be to
explore this idea to observe the perturbative phase in a  "macroscopic" volume.
One manifestation of this would be the production of huge
blob--like configurations of Huskyons
\cite{[10]}. These different configurations have wildly
different interactions with a nucleus \cite{[11]},
so that the nucleon in the nucleus can be
very different from a free nucleon.   More generally,  the idea
that a nucleon is a composite object is emphasized by these findings.
Some configurations of the nucleon interact very strongly with the surrounding
nucleons; some interact very weakly.   This means that
the nucleon in the nucleus can be very different from a free nucleon.
This leads to an entirely new view of the nucleus, one in which
the nucleus is made out of oscillating, pulsating,
vibrating, color singlet, composite objects.

The technical purpose of this paper has been to
show how to apply  leading-order
perturbative QCD to computing the scattering
amplitude for the coherent  processes:
$\pi N\to JJ\; N$ and $\pi A \to JJ\; A$.
The high momentum component of the  pion wave function, computable
in perturbation theory, is an essential element of
the amplitude. The dominance of the amplitude of the $T_1$ term of
Eq.~(\ref{t1r}) is obtained by showing  that the corrections to it,
which at first glance seem to be of the same order in
the coupling constant, are vanishingly small.
This vanishing, obtained using arguments based on analyticity, causality,
and current conservation, is
equivalent to the verification of a
specific space-time description of the
event:    the pion produces its point-like
component at distances well before the target. Furthermore, for the
conditions of the experiment \cite{danny}
studied here, the competing  electromagnetic
production process is shown to yield a negligible effect.
It seems  that perturbative QCD can be applied
to the coherent nuclear  production
of high-relative momentum dijets by high energy pions.

It therefore seems interesting to consider similar reactions involving
other projectiles such as photons, kaons, and protons.
The observations of the coherent photoproduction  of the $J/\psi$
from nuclear targets has long been known\cite{Sokoloff} to have an
A dependence which is very similar  to that observed here, but the
authors of \cite{Sokoloff} did not interpret it as color
transparency phenomenon. Later on H1 and ZEUS detectors at HERA
investigated exclusive photoproduction of the $J/\psi $ meson from a
proton target. The  striking qualitative predictions for this process
based on the QCD factorization theorem,  such as energy and  t dependence,
are in accordance with the data; for a recent review see \cite{AC,MWUST}.
Thus now there exist serious reasons to believe that
color transparency phenomenon has been observed in the combination of coherent
photoproduction of $J/\psi$ from  nuclei(FNAL) and from  the nucleon (HERA).
Our present theory can be used for kaon projectiles with little modification.
Because  the kaon has  a smaller size than the pion, we expect that the
amplitude  for a kaon-induced process should be somewhat larger than that
of the pion induced process discussed here. The same analysis should be
applicable to the photoproduction of high $\kappa_t$  $q\bar q $ pair of
light quarks. It is important here that the contribution of
bare photon coupling to  the $q\bar q$ 
 is proportional to quark's bare mass  and is
negligible for forward scattering \cite{brodsky,diehl}. The
contribution of a target gluon with $k_{it}^2\approx \kappa_t^2$
discussed in \cite{diehl} is suppressed by Sudakov and $w_2$ form
factors discussed above.  However, the PQCD physics of light quarks
will be masked to some extent by another striking prediction of QCD,
which is the enhancement of the diffractive production of charmed
dijets because of a large bare mass of charmed quarks.

The study of high energy coherent  production of jet
systems from nuclear targets seems to be a very productive way to investigate
both   perturbative QCD and microscopic
nuclear structure by exploring the diverse effects of color transparency.
Such studies in the region of hundreds of GeV
seem  ideally suited for the non-destructive investigation
of a microscopic hadron, and nuclear structure. It
may be possible to remove a piece of hadron
($q\bar q$ pair..) or to implant (strangeness in the center of
a nucleus...) without destruction of a target.
Such investigations resemble modern methods of surgery which avoid
cutting muscles. So it seems appropriate to name this new field
of investigations as micro-surgery of a hadron, or  of a nucleus.

\section*{Acknowledgments}
We would like to thank S.Brodsky, J.Collins, Yu.Dokshitzer and
A.Mueller for the useful discussions.
This work has been supported in part by the USDOE and GIF.


\begin{references}
\bibitem{AC}
 H.~Abramowicz and A.~Caldwell, Rev. Mod. Phys. {\bf 71}, 1275 (1999).

\bibitem{Randa}S.~F.~King, A.~Donnachie and J.~Randa,
%``Diffractive Photoproduction Of Quark - Anti-Quark Jets,''
Nucl.\ Phys.\  {\bf B167}, 98 (1980);


J.~Randa,
%``Quark Jets In Pion Diffractive Dissociation,''
Phys.\ Rev.\  {\bf D22}, 1583 (1980).



\bibitem{bb}G.F.~Bertsch, S.J.~Brodsky, A.S.~Goldhaber, and J.F.~Gunion,
%``Diffractive Excitation In QCD,"
Phys. Rev. Lett. {\bf 47}, 297 (1981).
%%CITATION = PRLTA,47,297;%%
%%%%%%%%%%%%%%%%%%%%%%%%%%%%%%%%%%%%%%%%%%%%%%%%%%%%%lf
\bibitem{comment}
 Reference ~\cite {bb} stated that
``{\it The exponential suppression we obtained in eq.(13) is of course
a many-body effect, with  validity limited to nuclear targets
and low to moderate $k_\perp$.
In the regime where the jet structure is
visible, the pion induced jets will have the angular dependence in the jet
c.m. frame of $d\sigma /dM^2 dcos \theta \approx sin^2\theta$, reflecting
the x dependence of}''  the equation for the pion wave
function. Thus  Ref.~\cite {bb} implied that $k_t$ dependence will obey a
power
law form: $k_\perp^{-6}$ of \cite{Randa},  which is different from the form
$k_\perp^{-8}$
deduced in this paper.
The conclusion   Ref.~\cite {bb} explicitly formulated is that
``{\it the nuclear filter does not generally produce states with large enough
$k_\perp$ that a jet structure is expected to appear.}''



\bibitem{fms93}
L. Frankfurt, G.A. Miller, and M.Strikman, Phys. Lett. {\bf B304}
  1, (1993).



\bibitem{Gribov}
V.~N.~Gribov,
``Lectures On The Theory Of Complex Momenta,''
KHFTI-PREPRINT-70-47; V.~N.~Gribov, L.~N.~Lipatov, and G.~V.~Frolov,
Sov.\ J.\ Nucl.\ Phys.\  {\bf 12}, 543 (1971).


\bibitem{Feynman}Feynman, R.P.
   PHOTON - HADRON INTERACTIONS.  By R.P. Feynman. Benjamin,
   1972. 
\bibitem{Gribov-Lipatov}V.~N.~Gribov and L.~N.~Lipatov,
%``Deep Inelastic E P Scattering In Perturbation Theory,''
Yad.\ Fiz.\  {\bf 15}, 781 (1972)
[Sov.\ J.\ Nucl.\ Phys.\  {\bf 15}, 438 (1972)].
\bibitem{Dok}Y.~L.~Dokshitzer,
%``Calculation Of The Structure Functions For Deep Inelastic Scattering
%And E+ E- Annihilation By Perturbation Theory In Quantum Chromodynamics.
%(In Russian),''
Sov.\ Phys.\ JETP {\bf 46}, 641 (1977).
\bibitem{BFKL}
E.~A.~Kuraev, L.~N.~Lipatov, and V.~S.~Fadin,
%``The Pomeranchuk Singularity In Nonabelian Gauge Theories,''
Sov.\ Phys.\ JETP {\bf 45}, 199 (1977)
[Zh.\ Eksp.\ Teor.\ Fiz.\  {\bf 72}, 377 (1977)].

I.~I.~Balitsky and L.~N.~Lipatov,
%``The Pomeranchuk Singularity In Quantum Chromodynamics,''
Sov.\ J.\ Nucl.\ Phys.\  {\bf 28}, 822 (1978)
[Yad.\ Fiz.\  {\bf 28}, 1597 (1978)].

\bibitem{FFGS}
L.~Frankfurt, A.~Freund, V.~Guzey, and M.~Strikman,
%``Nondiagonal parton distributions in the leading logarithmic  approximation,''
Phys.\ Lett.\ B {\bf 418}, 345 (1998)
[Erratum-ibid.\ B {\bf 429}, 414 (1998)]
[hep-ph/9703449].

\bibitem{fmsrev} See the review:
L.L.~Frankfurt, G.A.~Miller and M.~Strikman,
%``The Geometrical color optics of coherent high-energy processes,"
Ann. Rev. Nucl. Part. Sci. {\bf 44}, 501 (1994)
hep-ph/9407274.


\bibitem{flow1}
F.E.~Low,
%``A Model Of The Bare Pomeron,''
Phys.\ Rev.\ {\bf D12}, 163 (1975).


\bibitem{danny}
E.~M.~Aitala {\it et al.}  [Fermilab E791 Collaboration],
%``Observation of color-transparency in diffractive dissociation of pions,''
Phys.\ Rev.\ Lett.\  {\bf 86}, 4773 (2001);
\noindent
E.~M.~Aitala {\it et al.}  [Fermilab E791 Collaboration],
Phys.\ Rev.\ Lett.\  {\bf 86}, 4768 (2001).

\bibitem{explain} Is it interesting to  note
that soft diffraction cross sections vary as $ A^{0.8}$, which is a higher
power  than the $\sim A^{1/3}$  expected from edge effects
which one would expect for  the case of scattering
from  a black body. A faster $A$ dependence arises from the
significant fluctuations in the strength of pion-nucleon interaction
(for a review and references see \cite{fmsrev}).

%gm hep-ex/0010043.



\bibitem{cfs} J.~C.~Collins, L.~Frankfurt and M.~Strikman,
%``Factorization for hard exclusive electroproduction of mesons in QCD,''
Phys.\ Rev.\  {\bf D56}, 2982 (1997)
[hep-ph/9611433].
\bibitem{hallertext} L.~Frankfurt, G.~A.~Miller and M.~Strikman,
%``Perturbative pion wave function in coherent pion nucleon
%dijet  production,''
Found.\ Phys.\  {\bf 30}, 533 (2000)





\bibitem{bbfs93}
B.~Blattel, G.~Baym, L.~L.~Frankfurt and M.~Strikman,
%``How transparent are hadrons to pions?,''
Phys.\ Rev.\ Lett.\  {\bf 70}, 896 (1993).
\bibitem{guzey}
L.~Frankfurt, V.~Guzey and M.~Strikman,
%``Color coherent phenomena on nuclei and the QCD evolution equation,''
J.\ Phys.\ G {\bf 27}, R23 (2001)
[arXiv:hep-ph/0010248].


\bibitem{FLS90}L.~L.~Frankfurt, M.~I.~Strikman and S.~Liuti,
%``Evidence For Enhancement Of Gluon And Valence Quark Distributions In
%Nuclei From Hard Lepton Nucleus Processes,''
Phys.\ Rev.\ Lett.\  {\bf 65}, 1725 (1990).

\bibitem{Pirner}T.~Gousset and H.~J.~Pirner,
%``The Ratio of Gluon Distributions in Sn and C,''
Phys.\ Lett.\ B {\bf 375}, 349 (1996)
[hep-ph/9601242].
\bibitem{Eskola}K.~J.~Eskola, V.~J.~Kolhinen and C.~A.~Salgado,
%``The scale dependent nuclear effects in parton distributions for
%practical applications,''
Eur.\ Phys.\ J.\  {\bf C9}, 61 (1999)
[hep-ph/9807297].




\bibitem{fs99}
L.~Frankfurt and M.~Strikman,
%``Diffraction at HERA, color opacity and nuclear shadowing,''
Eur.\ Phys.\ J.\ {\bf A5}, 293 (1999).

\bibitem{NMC}M.~Arneodo {\it et al.}  [New Muon Collaboration],
%``The Q**2 dependence of the structure function ratio F2 Sn / F2 C and
%the difference R Sn - R C in deep inelastic muon scattering,''
Nucl.\ Phys.\ B {\bf 481}, 23 (1996).


\bibitem{fks} L. Frankfurt, W. Koepf, M. Strikman,
 Phys. Rev.  {\bf D54}, 3194 (1996);
M.~McDermott, L.~Frankfurt, V.~Guzey and M.~Strikman,
%``Unitarity and the QCD-improved dipole picture,''
hep-ph/9912547.
\bibitem{gv}
M.~Vanderhaeghen, P.~A.~Guichon and M.~Guidal,
%``Hard electroproduction of photons and mesons on the nucleon,''
Phys.\ Rev.\ Lett.\  {\bf 80}, 5064 (1998).



 \bibitem{real}
V.~N.~Gribov and A.~A.~Migdal,
%``Properties Of The Pomeranchuk Pole And The Branch Cuts Related
%To It At Low Momentum Transfer,''
Sov.\ J.\ Nucl.\ Phys.\  {\bf 8}, 583 (1969);

J.~B.~Bronzan, G.~L.~Kane and U.~P.~Sukhatme,
%``Obtaining Real Parts Of Scattering Amplitudes Directly From
%Cross-Section Data Using Derivative Analyticity Relations,''
Phys.\ Lett.\  {\bf B49}, 272 (1974).


   \bibitem{t'hooft}G.~'t Hooft,
%``Renormalization Of Massless Yang-Mills Fields,''
Nucl.\ Phys.\  {\bf B33}, 173 (1971);
%``Renormalizable Lagrangians For Massive Yang-Mills Fields,''
G.~'t Hooft, Nucl.\ Phys.\  {\bf B35}, 167 (1971);
G.~'t Hooft and M.~Veltman,
%``Regularization And Renormalization Of Gauge Fields,''
Nucl.\ Phys.\  {\bf B44}, 189 (1972).
\bibitem{Frankfurt:1991rk}
L.~Frankfurt and M.~Strikman,
%``Color screening and color transparency in hard nuclear processes,''
Prog.\ Part.\ Nucl.\ Phys.\ {\bf 27}, 135 (1991).
%%CITATION = PPNPD,27,135;%%

\bibitem{higgs}
  Y.~L.~Dokshitzer, V.~A.~Khoze, and S.~I.~Troyan,
in Proceedings of the 6$^{th}$ Int.
 Conference on Physics in Collisions  1986,
ed M. Derrick (World Scientific, Singapore, 1987), p. 365;

H.~Bengtsson and T.~Sjostrand,
%``The Lund Monte Carlo For Hadronic Processes: Pythia Version 4.8,''
Comput.\ Phys.\ Commun.\  {\bf 46}, 43 (1987);



Y.~L.~Dokshitzer, V.~A.~Khoze, S.~I.~Troian and A.~H.~Mueller,
%``QCD Coherence In High-Energy Reactions,''
Rev.\ Mod.\ Phys.\  {\bf 60}, 373 (1988).

\bibitem{brodsky}
S.~J.~Brodsky and J.~Gillespie,
%``Second Born Corrections To Wide Angle Electron Pair Production And
%Bremsstrahlung,''
Phys.\ Rev.\  {\bf 173}, 1011 (1968).

\bibitem{diehl}
M.~Diehl,
%``Diffractive production of dijets at HERA,''
Z.\ Phys.\ C {\bf 66}, 181 (1995)
[hep-ph/9407399].
\bibitem{WW} V.B.Berestetskii, E.M. Lifshitz, L.P. Pitaevskii
   Quantum Electrodynamics.  2nd ed.  Pergamon Press, 1982.
   652p.  (Course of Theoretical Physics, v. 4)
\bibitem{TTWu} Hung Cheng and Tai Tsun Wu,
 Expanding Protons: scattering at high energies,
MIT Press, 1987. 285p.

\bibitem{flow}
F.E.~Low,
Phys.\ Rev.\  {\bf 110}, 974 (1958).



\bibitem {bfgms}
  S.~J.~Brodsky, L.~Frankfurt, J.~F.~Gunion, A.~H.~Mueller and M.~Strikman,
%``Diffractive leptoproduction of vector mesons in QCD,''
Phys.\ Rev.\  {\bf D50}, 3134 (1994).


\bibitem{torder} Each Feynman graph shown in our figures consists of a sum of
  different graphs corresponding to different time orderings. When we define
a given intermediate state, we are concerned with a given time ordering, and
will find it necessary to consider other time orderings of the same
Feynman graph.
\bibitem{Braun}
V.~M.~Braun, D.~Y.~Ivanov, A.~Schafer and L.~Szymanowski,
%``QCD factorization for the pion diffractive dissociation to two jets,''
Phys.\ Lett.\ B {\bf 509}, 43 (2001).


\bibitem{IS}G.~Ingelman and P.~E.~Schlein,
%``Jet Structure In High Mass Diffractive Scattering,''
Phys.\ Lett.\ B {\bf 152}, 256 (1985).

\bibitem{H1}
C.~Adloff {\it et al.}  [H1 Collaboration],
%``Diffractive jet production in deep-inelastic e+ p collisions at HERA,''
Eur.\ Phys.\ J.\ C {\bf 20}, 29 (2001)
[arXiv:hep-ex/0012051].

\bibitem{CFS93}
J.~C.~Collins, L.~Frankfurt and M.~Strikman,
%``Diffractive hard scattering with a coherent pomeron,''
Phys.\ Lett.\ B {\bf 307}, 161 (1993)
[arXiv:hep-ph/9212212].

 \bibitem{jm} B.K. Jennings and G.A. Miller Phys.Rev. {\bf C50}, 3018 (1994).

\bibitem{sudff}
J.~M.~Cornwall and G.~Tiktopoulos,
%``Infrared Behavior Of Nonabelian Gauge Theories. 2,''
Phys.\ Rev.\ D {\bf 15}, 2937 (1977).

R.~Coquereaux and E.~de Rafael,
%``Large Transverse Momentum Behavior Of Gauge Theories,''
Phys.\ Lett.\ B {\bf 69}, 181 (1977).

\bibitem{DDT}
Y.~L.~Dokshitzer, D.~Diakonov and S.~I.~Troian,
%``Hard Processes In Quantum Chromodynamics,''
Phys.\ Rept.\  {\bf 58}, 269 (1980).


\bibitem{nnn}
N.~N.~Nikolaev, W.~Schafer and G.~Schwiete,
%``Coherent production of hard dijets on nuclei in QCD,''
hep-ph/0009038
Phys.\ Rev.\ D {\bf 63}, 014020 (2001).
\bibitem{BL} S.J. Brodsky G.P. Lepage,
Phys. Rev. {\bf D22}, 2157 (1982)
\bibitem {arrr} A.~V.~Efremov and A.~V.~Radyushkin,
%``Factorization And Asymptotical Behavior Of Pion Form-Factor In QCD,''
Phys.\ Lett.\ B {\bf 94}, 245 (1980).
%%CITATION = PHLTA,B94,245;%%
G.~P.~Lepage and S.~J.~Brodsky,
%``Exclusive Processes In Quantum Chromodynamics: Evolution
%Equations For Hadronic Wave Functions And The Form-Factors Of Mesons,''
Phys.\ Lett.\ B {\bf 87}, 359 (1979).


\bibitem{tolya}
A.V.~Radyushkin,
%``QCD sum rules: Form-factors and wave functions,"
hep-ph/9707335;
A.~Szczepaniak, A.~Radyushkin and C.~Ji,
%``Consistent analysis of O (alpha-s) corrections to pion elastic form
%                  factor,"
Phys. Rev. {\bf D57}, 2813 (1998)
hep-ph/9708237;
I.V.~Musatov and A.V.~Radyushkin,
%``Transverse momentum and Sudakov effects in exclusive QCD processes: Gamma*
%                  gamma pi0 form-factor,"
Phys. Rev. {\bf D56}, 2713 (1997)
hep-ph/9702443.


\bibitem{kroll} P.~Kroll and M.~Raulfs,
%``The Pi gamma transition form-factor and the pion wave function,"
Phys. Lett. {\bf B387}, 848 (1996)
hep-ph/9605264.




\bibitem{frs}L. Frankfurt, A. Radyushkin, and M. Strikman,
  Phys. Rev. {\bf D55}, 98 (1997)


\bibitem{Lech}
L.~Mankiewicz, G.~Piller and T.~Weigl,
%``Hard exclusive meson production and nonforward parton distributions,''
Eur.\ Phys.\ J.\ C {\bf 5}, 119 (1998)
[arXiv:hep-ph/9711227].
\bibitem{fg} A. Freund, V. Guzey,
 hep-ph - 9801388;
Phys.\ Lett.\ B {\bf 462}, 178 (1999)
[arXiv:hep-ph/9806267].

\bibitem{abramowicz}
H.~Abramowicz, L.~Frankfurt and M.~Strikman,
%``Interplay of hard and soft physics in small x deep inelastic processes,''
SLAC Summer Inst.1994:0539-574,
Surveys High Energy.\ Phys.\  {\bf 11}, 51 (1997)
[hep-ph/9503437].
\bibitem{radyushkin}
A.~V.~Radyushkin,
Phys.\ Rev.\  {\bf D59}, 014030 (1999)


\bibitem{BW} R.~H.~Bassel, and C.~Wilkin, Phys.Rev. {\bf 174} 1179 (1968).



\bibitem{glauberadep}
L.~Frankfurt, G.~A.~Miller and M.~Strikman,
%``Evidence for color fluctuations in hadrons from coherent
%nuclear diffraction,''
Phys.\ Rev.\ Lett.\  {\bf 71}, 2859 (1993)
[hep-ph/9309285].



\bibitem{Ferbel}
M.~Zielinski {\it et al.},
%``Three Pion Production On Nuclei At 200-Gev,''
Z.\ Phys.\  {\bf C16}, 197 (1983).


\bibitem{CFS}J.~C.~Collins, L.~Frankfurt and M.~Strikman,
%``Factorization for hard exclusive electroproduction of mesons in QCD,''
Phys.\ Rev.\ D {\bf 56}, 2982 (1997)
[hep-ph/9611433].


\bibitem{[7]} A.S.~Carroll {\it et al.},
%``Nuclear Transparency To Large Angle P P Elastic Scattering,''
Phys. Rev. Lett. {\bf 61}, 1698 (1988);



\bibitem{[8]}
B.K.~Jennings and G.A.~Miller,
%``Realistic hadronic matrix element approach to color transparency,''
Phys. Rev. Lett. {\bf 69}, 3619 (1992);



B.K.~Jennings and G.A.~Miller,
%``On Color Transparency,''
Phys. Lett. {\bf B236}, 209 (1990);



B.K.~Jennings and G.A.~Miller,
%``Energy dependence of Color Transparency''
Phys. Rev. {\bf D44}, 692 (1991);

B.K.~Jennings and G.A.~Miller,
%``Color transparency in (p, p p) reactions,''
Phys. Lett. {\bf B318}, 7 (1993).




\bibitem{[9]} K.~Egiyan,
     L.~Frankfurt, W.R.~Greenberg, G.A.~Miller, M.~Sargsian and M.~Strikman,
Nucl. Phys. {\bf A580}, 365 (1994);
L.L.~Frankfurt, W.R.~Greenberg, G.A.~Miller, M.M.~Sargsian and M.I.~Strikman,
Z. Phys. {\bf A352}, 97 (1995);
L.~Frankfurt, W.R.~Greenberg, G.A.~Miller, M.M.~Sargsian and M.I.~Strikman,
Phys. Lett. {\bf B369}, 201 (1996)
.%nucl-th/9412033.

\bibitem{IS1} 
D.~Y.~Ivanov and L.~Szymanowski,
%``Coulomb dissociation of a fast pion into two jets,''
Phys.\ Rev.\ D {\bf 64}, 097506 (2001)
[arXiv:hep-ph/0103184].




\bibitem{[10]} A.~Bulgac and L.~Frankfurt,
Phys. Rev. {\bf D48} (1993) 1894.
 \bibitem{[11]}
L.L.~Frankfurt and M.I.~Strikman,
Nucl. Phys. {\bf B250}, 143 (1985).
M.R.~Frank, B.K.~Jennings and G.A.~Miller,
Phys. Rev. {\bf C54}, 920 (1996).

\bibitem{Sokoloff}
M.~D.~Sokoloff {\it et al.}  [Fermilab Tagged Photon Spectrometer
                  Collaboration],
%``An Experimental Study Of The A-Dependence Of J / Psi Photoproduction,''
Phys.\ Rev.\ Lett.\  {\bf 57}, 3003 (1986).

\bibitem{MWUST}
M.~Wusthoff and A.~D.~Martin,
%``The QCD description of diffractive processes,''
J.\ Phys.\ {\bf G25}, R309 (1999)
[hep-ph/9909362].

\end{references}
\end{document}